\documentclass[letterpaper,twocolumn,10pt]{article}
\usepackage{usenix-2020-09}
\usepackage{censor}
\usepackage[available,functional]{usenixbadges}

\expandafter\def\expandafter\UrlBreaks\expandafter{\UrlBreaks
  \do\a\do\b\do\c\do\d\do\e\do\f\do\g\do\h\do\i\do\j%
  \do\k\do\l\do\m\do\n\do\o\do\p\do\q\do\r\do\s\do\t%
  \do\u\do\v\do\w\do\x\do\y\do\z\do\A\do\B\do\C\do\D%
  \do\E\do\F\do\G\do\H\do\I\do\J\do\K\do\L\do\M\do\N%
  \do\O\do\P\do\Q\do\R\do\S\do\T\do\U\do\V\do\W\do\X%
  \do\Y\do\Z}
  
\begin{document}

%don't want date printed
\date{}


\title{Web Execution Bundles:\\ Reproducible, Accurate, and Archivable Web Measurements}

% \author{Anonymous submission}

% %for single author (just remove % characters)
% \author{
% {Florian Hantke}\\
% CISPA Helmholtz Center for Information Security
% \and
% {Peter Snyder}\\
% Brave Software
% \and
% {\rm Hamed Haddadi}\\
% Imperial College London \& Brave Software
% \and
% {\rm Ben Stock}\\
% CISPA Helmholtz Center for Information Security
% } % end author

\author{
{\rm Florian Hantke${}^\dagger$, Peter Snyder${}^\ddagger$, Hamed Haddadi${}^\star{}^\ddagger$, Ben Stock${}^\dagger$} \\
\small{$\dagger$ CISPA Helmholtz Center for Information Security $\ddagger$ Brave Software $\star$ Imperial College London}\\
\small{ 
\href{mailto:florian.hantke@cispa.de}{
    \textcolor{black}{florian.hantke@cispa.de}
},
\href{mailto:pes@brave.com}{
    \textcolor{black}{pes@brave.com}
},
\href{mailto:h.haddadi@imperial.ac.uk}{
    \textcolor{black}{h.haddadi@imperial.ac.uk}
},
\href{mailto:stock@cispa.de}{
    \textcolor{black}{stock@cispa.de}
}
}}

\maketitle

%-------------------------------------------------------------------------------
\begin{abstract}
\revise{
% With a new topic this year, USENIX Security appropriately sets a focus on the reproducibility of research.
Recently, reproducibility has become a cornerstone in the security and privacy research community, including artifact evaluations~\cite{usenixCfP, ndssCfP, ccsCfP} and even a new symposium topic~\cite{usenixCfP}.}
However, \Web{} measurements lack tools that can be reused across many measurement tasks without modification, while being robust to circumvention, and accurate across the wide range of behaviors in the \Web{}.
As a result, most measurement studies use custom tools and varied archival formats, each of unknown correctness and significant limitations,
systematically affecting the research's accuracy and reproducibility. % of \Web{} measurements.

%Web measurement also lacks general archival formats that allow accurate measurement of past website behavior.
% \revise{and influence future research practices}
To address these limitations, we present \tool{}, a \Web{} measurement tool that is, compared against the current state-of-the-art, \emph{accurate} (\IE{} correctly measures and attributes events not possible with existing tools), \emph{general} (\IE{} reusable without modification for a broad range of measurement tasks), and \emph{comprehensive} (\IE{} handling events from all relevant browser behaviors). We also present \format{}, an archival format for the accurate and reproducible measurement of a wide range of website behaviors. We empirically evaluate \tool{}'s accuracy by replicating well-known \Web{} measurement
studies and showing that \tool{}'s results more accurately match our baseline. We then assess if \tool{} and \format{} succeed as general-purpose tools, which could be used to accomplish many \Web{} measurement tasks without modification. 
We find that this is so: 70\% of papers discussed in a 2024 web crawling SoK paper could be conducted using \tool{} as is, and a larger number (48\%) could be leveraged against \format{} archives without requiring any new crawling.

\end{abstract}
%-------------------------------------------------------------------------------

\section{Introduction}
\label{sec:intro}
Security and privacy \Web{} measurement is an enormous field that lacks accurate, general,
and standardized measurement tools and archival formats needed for reproducible measurements and general-purpose datasets. We argue that existing tools in the field operate at the wrong level of abstraction and granularity. 
In the best of cases, this mismatch invites error; in many cases, it makes certain categories of error unavoidable.

% the current tools and formats we have are X Y Z, and they are have problems like A B C~\cite{jueckstockRealisticReproducibleWebCrawl2021}.

As a demonstrative example of how the most common current \Web{} measurement tools
and \Web{} archive formats invite errors, consider the following example. Say a researcher
wanted to measure how the Web behaved in 2010 (the first year the HTTPArchive project~\cite{httparchive}
provides archives). \revise{The common way such measurements are done is to use
a browser automation framework like Puppeteer~\cite{puppeteer}}
% (an open source library used to instrument browsers for \Web{} measurement tasks) %described in Section \ref{sec:back:tools:instrumentation})
to automate a browser and load a website from a structured HTTP archive such as a HAR file.
 % (a structured archive of all HTTP requests and responses that occurred when loading a website). 
Current implementations of Puppeteer use the current version of Chromium by default and are incompatible with very old browser versions, meaning that websites archived in 2010
will be executed in a 2024 (or later) version browser. One large change that occurred in
the \Web{} platform between 2010 and 2024 is that browsers changed how they handle certain ``mixed 
content'' resources, or insecure sub-resources included on a secure website. Before 2015, Chromium
\emph{would} load insecure \JS{} included on a secure page, but (in
alignment with other browsers) after 2015, Chromium \emph{would not} load it on a secure page. The net effect of this change is that measuring how the \Web{} behaved
in 2010 by using modern Chromium versions 
% to load websites from 2010-archived HAR files
will cause several subtle but categorical measurement errors. It will
under-count \JS{} sub-resource requests while over-counting the number of failed HTTP requests,
among other errors. This is just one example.

Current measurement
tools can potentially cause many other errors of this kind, as well as other entirely different classes of errors. Using current archival formats for historical \Web{} measurements will cause comparable problems. More broadly, the lack of accurate, general-purpose \Web{} measurement tools and archive formats
weakens the field of \Web{} security and privacy measurement in at least the following ways:
\textit{i.} by encouraging, or sometimes requiring, measurement errors; \textit{ii.} by making
it difficult to compare results across measurement studies (since differing results
could be introduced by changes in measurement tools, measurement vantage point, etc.); \textit{iii.} by preventing accurate and reproducible historical measurements; and
\textit{iv.} by consuming an enormous amount of researcher time on
developing similar-but-bespoke measurement tools for each task.

%Further, we note that the problem of lacking general purpose, reusable, and accurate measurement tools (along with the related problem of lacking reusable, accurate, general purpose datasets) is relatively unique among other areas of computer science research. Fields like machine learning, artificial intelligence, natural language processing, and system performance  (incomplete list) have common datasets that are reused across studies, papers, and systems.   % HH: embedded this into the next paragraph

Research areas benefit greatly from having common, reusable datasets. For example, such datasets
encourage incremental improvements,
% (instead of ``reinventing the wheel'' for each paper),
and allow the results to be easily, objectively, and accurately compared across papers.
% (instead of results that are specific to particular measurement tools and per-study datasets).
A primary reason why
the field of security and privacy \Web{} measurement lacks such datasets is that the field
lacks a general-use, accurate format for archiving \Web{} measurements, which would allow a common
dataset to be reused across measurement tasks. By comparison, such datasets exist in other areas of computer science research, such as machine learning and system performance.

In this work, we aim to improve the state of security and privacy measurement in two ways.
First, by presenting \tool{}, a system for accurately measuring how browsers execute websites,
in a manner robust to page circumvention, and covering a broad enough range of browser
behaviors (\EG{} rendering, network, script execution, etc.)
that \tool{} can be used without modification for extensive measurements. And
second, with \format{}, a novel archival format for recording how a browser fetched, rendered,
and executed a website.

\tool{} and \format{} bring the following benefits for
precise and reproducible \Web{} measurement. First, they provide \textbf{comprehensiveness} where current tools do not. \tool{} is deeply integrated with Chromium and is built 
with industry-assisted knowledge of browser architecture to ensure both comprehensive and
correct results (see \Cref{sec:back:tools:modification}; this differs significantly
from current Web measurement tools).

Second, \tool{} and \format{} enable \textbf{accurate} archiving. \tool{} records both the 
resources that were captured during page execution, along with a fine-grained execution timeline. \format{} archives enable researchers to ask a wide range of questions about how websites behaved at
the time-of-archiving in a way fundamentally not possible with most current tools.

Third, \tool{} and \format{} are \textbf{general} tools, and allow researchers to measure
a wide range of browser behaviors without requiring modifications or customization.
By aiming to be general and reusable, \tool{} benefits from iterative improvements and corrections,
in a way uncommon to most measurement tools (which, historically, tend to be single-use and discarded
after each project).

More broadly, in this work, we make the following contributions to the goal of accurate and reproducible security and privacy \Web{} measurements:

\begin{compactenum}
    \item a \textbf{comprehensive overview of the \Web{} measurement and archiving tools} commonly used
        in research, along with the limitations why each is insufficient for the goal of generating accurate, reusable, achievable \Web{} data;
    \item the \textbf{design of \tool{} and \format{}}, a reusable tool and format for measuring and archiving how browsers execute websites accurately and robustly to allow reproducible measurements beyond what current tools and archival formats provide;
    \item multiple \textbf{empirical evaluations of \tool{}'s and \format{}'s accuracy}, compared
        against existing commonly used tools, finding that \tool{} more accurately matches baseline counts of events;
    \item an \textbf{empirical measurement of \tool's generality}, by surveying recent
        security and privacy \Web{} measurement papers and evaluating which could have been
        performed with \tool{} directly, or \format{} archives, avoiding the
        need to create new, custom measurement tools;
    \item the \textbf{open source implementation} of both, as modifications to Chromium (see~\Cref{apx:availability}) and related tools for crawling with \tool{}, generating \format{} archives, and querying them regarding security and privacy-relevant phenomena (\EG{} to query what Web APIs were called in which frames, among many other possibilities);
        % the request chains that occurred during page execution
 \end{compactenum}

Finally, we plan a \textbf{publicly available archive} in the \format{} format, enabling researchers to conduct a wide range of accurate,
replicable measurements of historical \Web{} behavior. We pledge to regularly update this archive
to create a common resource for \Web{} research comparable to the HTTPArchive.
       % a \textbf{publicly available archive} of \NumArchives{} websites in the \format{} format,
      %  to enable researchers to more conduct a wide range of accurate, replicable measurements of historical \Web{} behavior. We pledge to regularly update this collection of archives,
       % to create a common resource for \Web{} research, in a manner comparable to the HTTPArchive.

\section{Tools \& Techniques for Web Measurements}
\label{sec:back}
In this section, we provide a brief overview of existing tools, techniques, and
formats commonly used in \Web{} security and privacy measurements, based on a literature review of top-tier papers, \rev{such as those listed in crawling SoK and survey papers~\cite{stafeevSoKStateKrawlers2024, ahmadApophaniesEpiphaniesHow2020}}. Our
goal is to be demonstrative of the
common approaches used in research and discuss why these
are insufficient for the replication, accuracy, and iterative needs of
scientific research. We then follow by describing common formats used
for archiving websites, so that measurements can be done retrospectively, after
websites have been changed or are no longer available on the \Web{}. We highlight each approach's strengths, as well as its limitations
that make it insufficient as an accurate, general-purpose archiving format.

\subsection{The Web and Browsers}
\rev{
Before discussing tools and formats for \Web{} measurements, we briefly outline how the \Web{} and browsers function.

Users typically control a client, such as a browser or curl, to request a resource from a \Web{} server via HTTP requests. The server processes these and returns an HTTP response containing the requested resource and further metadata~\cite{mdnWeb}.

When browsers retrieve an HTML document, they parse it based on a well-defined specification~\cite{htmlSpecParsing}. This process involves tokenizing the input stream and constructing the Document Object Model (DOM) from it -- a tree-like representation of the page's content. CSS is parsed into the CSS Object Model, combined with the DOM to build the Render Tree. From this tree, the browser calculates element positions and dimensions before painting them on the screen.

As the DOM is processed, additional resources may be loaded and  JavaScript (JS) executions can modify the DOM. These resources and dynamic changes are not part of the initial HTTP response and require rendering and further requests. Once all resources are loaded, the \JS{} is executed, and the Render Tree is finalized and painted, the user sees the \Web{} page as the final product~\cite{mdnParsing}. To measure and analyze these behaviors for research, tools and archives are needed that accurately interpreting these standardized processes.
}

\subsection{Existing Web Measurement Tools}
\label{sec:back:tools}
We first describe four general categories of tools commonly used in \Web{} measurement.
\rev{Based on our literature research, we find that all \Web{} measurement tools used in top security and privacy papers fall into one of these categories.}

% We assert that either all, or nearly all, of the \Web{} measurement tools used in top security or privacy papers fall into one of these. 

%We highlight the strengths of each approach, along with each category's limitations which make them insufficient to be a general, accurate, and reusable approach for \Web{} measurement.       % HH: this sentence was repeated above

\subsubsection{Recording HTTP Requests}
\label{sec:back:tools:requests}
The simplest, and earliest, approach used in \Web{} measurement research is to
record the HTTP communication for each page
being measured. In this approach, researchers select a tool for issuing an HTTP
request to a URL (\EG{} curl~\cite{curl},
wget~\cite{wget}, among many
others), use the tool to request the HTML document for the webpage being
measured, and record the outgoing HTTP request and the server's HTTP response. \rev{Numerous examples can be found in literature~\cite{bollinger2022automating, calzavara2020tale, robertsYouAreWho2019, invernizziCloakVisibilityDetecting2016}.}

This approach can be extended to use the same tools to record the HTTP communication for sub-resources (\EG{} images, or script files)
referenced in the HTML of the webpage, either by searching for URL patterns or parsing the HTML.
% either by searching for URL patterns in
% the HTML text, or by parsing the HTML text into a structured representation, and
% then extracting the URLs for the relevant sub-resources.

The defining feature of this approach is that tools are
being used to understand or approximate how users experience the \Web{}, but
without actually incorporating the main tool people use to interact with the
\Web{}; \Web{} browsers.

\subsubsubsection{Limitations}
Measuring the \Web{} based on raw HTTP request data has important benefits;
hundreds of instances of \ttt{curl} can be run in the amount of
resources needed to load a single webpage in Chrome. However, this
approach also comes with serious limitations; Evaluating HTTP and
HTML without using a \Web{} browser will give, an extremely
\tit{lossy} approximation of peoples' \Web{} experience, making it unsuited for answering many \Web{} related research questions.

% It takes an extreme amount of domain-specific expertise to approximate \Web{} browser behavior when parsing HTML without using a \Web{} browser.

As example, \Web{} browsers use complicated rules for deciding which
sub-resources to fetch, affected by many factors like a page's \tit{Content Security Policy}~\cite{csp}, \ttt{preload}~\cite{htmlpreload} instructions given on previous pages, among others. \rev{Concretely, consider a dynamically added image element; the image is typically fetched when the element's \texttt{src} attribute is set, regardless of whether it has been appended to the DOM.} These edge cases require an extreme amount of domain-specific expertise\rev{, leading to accidental but \emph{preventable} errors}. 

In short, tools in this category focus on HTTP communications but do not consider how \Web{} browsers parse and execute
those websites.
% and sub-resources described by those HTTP responses.
Therefore, they are only suited to answer questions about server
responses, but not \rev{accurate enough} to measure how the \Web{} is experienced by most people.

\subsubsection{Browser Instrumentation}
\label{sec:back:tools:instrumentation}
A second approach is to leverage the browser itself, either through automation APIs (\IE{} WebDriver~\cite{webdriver} and
DevTools~\cite{devtoolsprotocol}) and
popular libraries for interacting with these APIs (\EG{}
Puppeteer~\cite{puppeteer}),
or through browser extensions. These approaches are used in a variety of \Web{} research~\cite{pletinckx2021out, sanchez2021journey, degeling2018we, rautenstrauchLeakyWebAutomated2023} and account for the limitations of
the previously discussed approach by measuring websites as executed by browsers.
%, instead of trying to approximate the browser environment with other tools.

All of these browser instrumentation techniques are able to measure how a
website is rendered, cookies and other values are stored, and execute and inspect \JS{} in the page
% to measure some aspects of page \JS{} use, 
by injecting proxies into global \JS{} structures.
% (either prototype chains or singleton instances).
Some tools also measure network requests made by the page.

\subsubsubsection{Limitations}
While these approaches are extremely common in \Web{} measurement research,
they have important limitations, making them unsuited for
measuring certain phenomena researchers investigate. Broadly speaking,
browser instrumentation tools lack APIs to directly
measure many page behaviors, requiring researchers to
% (again, as with the previously discussed approach)
try and write code that replicates
browser behavior. In the best of cases, this invites \emph{preventable} implementation errors; in
many worst cases, measurement errors are unavoidable or \emph{unpreventable}.

% For example, in some cases researchers are effectively measuring how the
% browsers \emph{should} work, which may be very different from how browsers
% actually work. The \Web{} platform is enormously complex, full of standardized
% exceptions, oddities, and inherited unexpected behaviors, which can result in a
% large gap between measurements and actual behavior\footnote{If you are skeptical
% of this claim, we invite the reader to to compare their mental model of complex
% browser behaviors like page navigation, frame loading, or speculative execution
% and element fetching (including error handling, event ordering, incomplete
% responses) and compare against the relevant standard documentation.}.

% Similarly,
\rev{For example,} browser-instrumentation-based measurement tools are not able to
observe \JS{} execution directly, requiring researchers to rely on
indirect best-effort techniques (\EG{} modifying the execution environment).
% by interposing on target APIs and properties).
These techniques lead to incomplete or incorrect results for a range of
reasons. Measurements that rely on modifying a script's execution
environment fail when page scripts regain access to unmodified \JS{}
prototypes \rev{(\EG{} by storing a reference to the original prototype early during page execution)}, something extremely difficult to prevent
% with out breaking or otherwise interfering with the page behaviors intended to be measured
% \footnote{As one example, absent specific protections, pages can
% regain access to unmodified prototypes by creating a local (\EG{}
% ``about:blank'' or similar) \ttt{<iframe>}, inserting it into the document, and
% extracting references to unmodified prototypes from the child frame's
% \ttt{.contentWindow} property. The countermeasures needed to protect
% against such prototype-recapture techniques can break or otherwise modify how an instrumented page behaves.}
and a pattern utilized by real-world
advertising libraries or anti-debugging techniques~\cite{musch2021u}. Furthermore, measurements relying on interposing of non-configurable global
\JS{} elements cannot capture the interaction with these (read-only) Web API properties, \rev{like \ttt{window.location}, %or \ttt{document.location},
making errors \emph{unpreventable} and leading to a lack of \emph{comprehensiveness} with current tools}.

These are just some examples of phenomena that browser-instrumentation-based
measurement tools have difficulties in \emph{accurately} measuring. 
% Other examples include attribution (\IE{} what actor caused an event to
% occur on the page), error handling and recovery, and behaviors that occur during
% parsing (both when parsing the initial page HTML, or when scripts access the
% parser via \ttt{.innerHtml}, \ttt{DOMParser}, etc.).
% While suitable for some measurement tasks, these tools also have significant limitations 
While suitable for functional tests (the task these tools were initially designed for~\cite{selenium-history}), these tools have significant limitations preventing them from being the \emph{general} purpose solutions in \Web{} measurement research.
% they are used for

\subsubsection{Browser Orchestration}
\label{sec:back:tools:orchestration}
A third category of \Web{} measurement tools builds on browser-instrumentation
tools, relying on the same browser capabilities and intervention points but
creating reusable and improvable instrumentation code for consistent measurement tasks.
This allows a common body of
code to improve over time by benefiting from the collective knowledge of
multiple researchers. In short, this third category of
\Web{} measurement tools improves measurement accuracy by pushing the level of
abstract up another layer, allowing researchers to conduct measurements more easily and
accurately without needing to personally understand the peculiarities of the \Web{} platform.

The most popular example of this category of \Web{} measurement tool is
\tit{OpenWPM}~\cite{englehardt2016online}, used by studies all across the community~\cite{bollinger2022automating, iqbal2022khaleesi, calzavara2021reining, zhang2021harpo}. In addition to the main benefit of
providing an open and reusable code base that researchers can use and
improve, \tit{OpenWPM} also has additional benefits. First, \tit{OpenWPM} provides
standardized inputs
% (\EG{} the URLs and Web APIs to be measured) 
and outputs
% (\EG{} a standardized database schema and format for resulting)
for conducting
measurements, further aiding replication.
% and future analysis.
And second,
it also includes scheduling functionality, making it easier to measure
websites in parallel across different machines.

\subsubsubsection{Limitations}
\tit{OpenWPM} relies on the same underlying systems and capabilities as the
previously discussed browser-instrumentation-based tools and so shares many of
the same limitations. While \tit{OpenWPM's} reusable, iteratively-improved
code base reduces the chances that researchers will make \emph{preventable errors}, \tit{OpenWPM} and similar systems fundamentally cannot address the \emph{unpreventable errors} caused by the limitations of the underlying capabilities OpenWPM relies on (see limitations in \Cref{sec:back:tools:instrumentation}), \rev{\EG{} not \emph{accurate} enough to \emph{comprehensively} measure all Web API properties.}

\subsubsection{Browser Modification}
\label{sec:back:tools:modification}
A fourth approach to \Web{} measurement is to try and avoid the limitations of the APIs
available to \emph{browser instrumentation} approaches by directly adding instrumentation to the browser's code base, \rev{for example the DOM construction}. Since all popular browsers are (in part or
in full) open source\rev{~\cite{chromiumSource, geckoSource, webkitSource}}, researchers can accurately
measure any browser functionality by modifying their code.

This browser-modification approach has been used for a wide range of measurement tasks,
generally to measure lower-level browser behaviors that are not directly
visible to page-executed \JS{} or the APIs available to the Puppeteer libraries.
Researchers have published research depending on browser modifications to
log script compilation and fine-grained \JS{} execution~\cite{vv8-imc19}, detect XSS attacks~\cite{stock2015facepalm}, the behavior of injected
style sheets~\cite{laperdrix2021fingerprinting}, and to perform taint analysis in \JS{}
code~\cite{kleinHandSanitizersWild2022},
% to measure the behavior of client-side sanitizing libraries
among many other examples.

\subsubsubsection{Limitations}
Measurement tools that modify how browsers are implemented provide researchers with
powerful measurement capabilities. However, this flexibility imposes
a wide range of limitations on these tools.
Primarily, the correct implementation of such tools requires an extremely high
amount of domain knowledge in the implementation and esoteric standardized behaviors of modern browsers, making \rev{\emph{preventable}} errors easy when instrumenting target behaviors. And while some of these errors may be obvious to researchers, \EG{} build errors, other errors
can be extremely subtle, and require domain expertise to catch, \EG{} miss-attribution of edge cases.
% (\EG{} miss-attribution
% of events, omitting uncommon or unknown causes of events from measurement results, crashes or
% result errors that occur rarely and/or non-deterministically).

As a demonstrative example of the difficulties of correctly implementing browser-modification-based measurement tools,
consider the task of modifying Chromium to log access to browser APIs (the goal of several projects, such as
\cite{iqbal2020adgraph,vv8-imc19}). Measuring which script is calling which
API \emph{completely and correctly} requires understanding i. the low-level implementation details,
(\EG{} the differences between a V8 \ttt{Isolate}, a V8 \ttt{Context}, a V8 \ttt{ScriptState}, and a blink \ttt{ExecutionContext})
ii. subtleties in the browser application model
(\EG{} the difference between the calling and receiving execution contexts),
and iii. a comprehensive knowledge of all code interacting with these details. This task is not uniquely difficult,
% among \Web{} measurement tasks,
and comparable difficulties exist for instrumenting the parsing behaviors, \JS{} ordering,
among others.

We emphasize that these \emph{are not} criticisms of existing research tools,
which often aim for a general understanding of one phenomenon rather than
complete correctness. Instead, we note these difficulties to show how
current tools are not well suited for the tasks of a \emph{generalized} \Web{} measurement tools, \rev{\IE{} they are not reusable for everyone for a wide range of measurement tasks without modifications}.

% \subsubsection{Relationship to \PG{}}
% \PG{} belongs to this last, ``browser
% modification'' category of \Web{} measurement tools.  However, \PG{}
% \emph{differs} from existing tools (both ``Visible V8'' and other proposed,
% browser-modifying measurement tools) in several important ways: generality (the
% ability to answer a diverse and wide range of measurement questions with a
% single, reusable code base), correctness (\PG{} is built with a deep
% understanding of browser standards and internals, and correctly handles
% categories of cases that existing tools either handle incorrectly, or do not
% handle at all), and coverage (\PG{} records a broad range of page events, along
% with the cause of each event, including network requests, document
% modifications, \JS{} compilation and calls, and cross-frame behaviors, among
% others).

\subsection{Existing Web Archiving Formats}
\label{sec:back:formats}
We next provide a brief overview of common archiving formats used in
\Web{} measurement.
% The following two categories describe commonly used formats in
% research and industry highlight the limitations of real world systems.

\subsubsection{URLs and Raw Resources}
\label{sec:back:formats:raw}
The simplest archive form generated by \Web{} crawls is
a list of visited pages' URLs, allowing revisits and re-measurements in the future (\EG{} the Tranco list~\cite{pochat2018tranco}). Lists like this are used to analyze phenomena like typosquatting~\cite{szurdi2014long} or similar.
% a measurement paper might include a dataset of URLs that
% included a particular \JS{} library, or that had a misconfigured security
% policy, with the intent of enabling future researchers to revisit the
% referenced websites and reproduce the reported results.

Other archival datasets might include the actual resources described
by the URL at measurement time, such as 
an image dataset containing both the URLs and the image's content.
Including both
makes the dataset more robust, since, even if
the hosting server no longer returns the resource from
the URL, the resource of interest is still available in the dataset.

\subsubsubsection{Limitations}
Such datasets are popular because they are easy to generate and to use. However, this approach has profound limitations, making it poorly suited as a \rev{\emph{general}} approach to \Web{} archiving.
% However, this approach to archiving findings has deep limitations that make it poorly suited as a general approach to \Web{} archiving.
Most fundamentally, such datasets lack many other
resources and information present when the website was loaded and
rendered \rev{(\EG{} HTTP headers)},  limiting the ability to ask new questions about old data.
% (a common use for archival data).

\subsubsection{HTTP Archives with Metadata}
\label{sec:back:formats:archives}

The other common form of \Web{} archiving is to record the HTTP headers and bodies of all requests and responses during the page's execution, along with metadata like timestamps. Frequently used formats are the HTTP Archive format (HAR) and
the Web Archive format (WARC), which record largely the same kinds of information,
though with minor differences, such as WARC de-duplicates requests while HAR
does not.

Sites archived in these formats record the
communication while executing a target \Web{} page. Researchers use them to
replicate past measurements~\cite{hantke2023you}, or perform new ones on historical \Web{} versions~\cite{pletinckx2021out, roth2020complex, lernerInternetJonesRaiders2016, stock2017web}. They replay responses by loading the initial request and the loaded sub-requests 
from the archive instead of from the network. Effectively, researchers attempt to
re-run the \Web{} site from the archive to approximate how the site
operated at measurement time.

\subsubsubsection{Limitations}
These archives are very useful for measurements that study the resources loaded by a page, as sub-resources are archived exactly. While extremely common in \Web{} privacy and security research, these approaches, however, are significantly \rev{\emph{inaccurate}} when trying to replicate how websites behaved during
execution.
% measuring request dependency chains, attributing
% which script units made which modifications on a page or executed which Web APIs,
% how a page's content was dynamically modified during execution, among many others.

There are multiple reasons why these archive formats do not accurately
enable future researchers to replicate how archived websites behaved. For example,
the web platform involves many sources of non-determinism (\EG{} the current
time, random values, available system resources, etc.) which can
cause websites to behave differently on re-execution \rev{-- an \emph{unpreventable} issue when this behavior was not captured}. 
%, in ways large and small.

Similarly, changes in browsers and archiving tool behavior over time
further introduces errors and inaccuracies.
% , when using these archives for measurements.
Browsers' behavior at \tit{archiving} time may differ significantly
from how browsers behave at \tit{replay} time, introducing entire categories of potential inaccuracies. For example, consider the aforementioned change of the \tit{mixed content} policy. Modern browsers block insecure \JS{} files while browsers ten years ago had no such restrictions~\cite{mixedcontent}.
This change means replaying such archives in modern \Web{} browsers will cause categorical inaccuracies, making it seem like users historically encountered far more blocked requests than they actually did.

% how changes in browser
% \tit{mixed content} policy could bias studies of past website behavior.
% Modern browsers refuse to
% load insecure \JS{} files on secure pages (\IE{} \ttt{https})
% ~\cite{mixedcontent}.
% This has not always been the case, and for much of the \Web{}'s history
% browsers did not have any such restriction. This change means replaying
% such archives in modern \Web{} browsers, will cause categorical inaccuracies, making it seem like \Web{} users historically encountered a lot more blocked
% requests and missing \JS{} files then they actually did.

This change is just one example of how
such archives can lead to systematic errors in historical
\Web{} measurement. Differences in the \tit{archiving} and the
\tit{replaying} environments in any of the following
will cause measurement errors: differences in i. browser (\EG{} Firefox \VS{} Chrome), ii.
browser version, iii. operating system, iv. updates
in available Web APIs, v. \tit{headless} or \tit{headed} browser modes, and vi. browser
language and locale preferences, 
% and vii. current state of browser-maintained security and privacy lists, 
among many others.
\rev{These errors might be \emph{preventable} when the exact infrastructure is replicatable, in many cases it is not~\cite{demirReproducibilityReplicabilityWeb2022}.}

\subsection{Summary: Limitations and Requirements} % in Existing Systems
\label{sec:back:limits}
\rev{Finally, we note the overall challenges facing researchers
attempting to conduct accurate, replicable, shareable \Web{} research, given the limitations in existing \Web{} measurement and archiving tools, leading to requirements for a new tool.}

\subsubsubsection{Prohibitive Required Domain Expertise}
\label{sec:back:limits:expertise}
The vast majority of non-trivial measurements of \Web{} behaviors require
a large amount of domain expertise. \emph{Accurate} measurements
demand understanding and often re-implementing various browser
behavior, challenging even for the
industry teams implementing browsers, let alone academic \Web{} researchers.
% (who's expertise may be more targeted to theoretical or conceptual concerns).

In the best of cases, the expertise needed for
\emph{accurate} measurement means researchers must
familiarize themselves with esoteric browser concepts to
\emph{accurately} approximate how users experience the \Web{}.
% (\EG{} the difference between a calling and a receiving
% execution context; the rules for event ordering; etc.).
% and error handling; how the ``back/forward'' cache interacts with the four-part
% frame navigation process; etc.).
In the worst of cases, the high level of required expertise results in incorrect results.

\subsubsubsection{Redundancy and Correctness}
\label{sec:back:limits:redundancy}
The current \Web{} measurement field lacks tools that are
both \emph{general} for use across a wide range of measurements, and implemented in a manner that yields \emph{accurate} results, \IE{}
avoid aforementioned \rev{\emph{comprehensiveness}} limitations (\Cref{sec:back:tools:instrumentation}).
Instead, % \Web{} measurement
projects typically implement a new \emph{instrumented browser},
typically without correctness tests beyond checks by the researchers themselves. The systematic result of such redundancy
is a \emph{worst of all worlds} situation: a lot of time spent
(\IE{} research teams re-implementing similar tools) creating
tools of unknown correctness (\IE{} often no correctness verification).

\subsubsubsection{Limited Dataset Sharing}
\label{sec:back:limits:context}
Despite the enormous amount of \Web{} measurement work conducted and published, our field lacks \emph{general} datasets for \rev{\emph{comprehensive}} measurement tasks. The closest our field has are URL lists
% (\EG{} the Tranco list\cite{pochat2018tranco}, Alexa Top sites), 
or collections of archive files, both with discussed limitations. The lack of a general dataset of website behavior hinders iterative improvements, as seen in other fields with common datasets like AI or ML.

\subsubsubsection{Tool Requirements}
As a result, \Web{} measurements result in independent executions of (often) different websites in independently implemented measurement tools, making it difficult to confidently compare results across papers. % Differences in accuracy, attribution, etc. may just as often stem from different tooling rather than actual phenomena. The lack of a general dataset of website behavior also hinders iterative improvements seen in other fields with common datasets like AI or ML.

\rev{To address this, a new tool is needed, especially to handle the \emph{unpreventable} errors from \Web{} API restrictions and replicability problems with dynamic behavior. We developed the following requirements for such a tool. First, it must be \emph{comprehensive} and support a wide range of measurements out of the box, including \JS{} executions, without requiring specialized domain expertise. Second, it must be \emph{accurate}, also when handling \JS{} observations.
% To enable these modifications, it should leverage a widely adopted browser that makes long-term maintenance straightforward and ensures correctness and validation of browser behavior.
And third, the tool must be \emph{general}, producing reusable archives that go beyond only storing a site's resources but also provide all the executions to enable accurate replication of experiments.
}
\section{\tool{} and \format{} Archives}
\label{sec:tool}
In this section, we introduce \textbf{Web Execution Bundle RECorder} (\tool{}), a tool for accurately measuring a wide range of
browser behaviors during the execution of a webpage, and \format{}, a three-part
archive bundle format that \tool{} uses for recording its observations.
The following subsections detail
the designs of \tool{} and \format{} and how they overcome the limitations of existing tools (\Cref{sec:back}). % , described in Section \ref{sec:back}.

\subsection{Overview and Goals}
\label{sec:tool:overview}
\tool{} is designed \rev{to be \emph{accurate}, \emph{general}, and \emph{comprehensive}, \IE{} it can} capture a wide range
of browser behaviors and events, making it suitable for a large fraction of security and privacy \Web{} measurements without the need for modification; we expect that data necessary for diverse studies is a subset of what \tool{} currently measures.

\tool{} records its measurements as a single \format{} archive, which is a bundle
of three child files, described in more detail in the following subsections. The format
of these files is consistent and well structured so that common tooling can be used to
query the page behaviors and events recorded in each \format{} archive, the same way that
SQL databases describing widely differing events can be queried using common tooling.

We note that \format{} archives contain all observed events that occurred
during \tool{}'s execution of the page and not a subset of behaviors specified at
\emph{measurement time}. This has the cost of making \format{} larger than needed for the immediate measurement task
but brings two corresponding benefits.

The first benefit of \tool{} producing \rev{\emph{comprehensive}} recordings
of page behavior is that \format{} archives are ideal for long-term, cross-purpose
archiving. These rich and comprehensive archives allow researchers
to ask a wide range of questions about historical data, including questions
that might have been unanticipated at \emph{archiving time}. We note that this makes \format{}
archives very different from the existing archive formats discussed in
Section \ref{sec:back:formats}, where datasets are generally narrowly focused on a
particular research question.

And second, \format{} is \rev{\emph{general}, allowing it to generate} archives to support reproducible studies beyond what existing common measurement tools and archive formats allow. \revise{Consider the demonstrative ``mixed content'' example from earlier, for instance}. Because \tool{}
directly records how the browser behaved when executing a page, it allows future researchers to
accurately measure how the website \revise{(including mixed content)} was experienced by past users. This is
very different from existing archive formats, which record \emph{what the browser fetched}
instead of \emph{how the browser behaved}. Only knowing what resources were fetched
for past websites hinders historical accuracy and reproducibility by making it
difficult-to-impossible to know if differing results are due to \emph{endogenous differences in
measurement correctness} (\EG{} detecting an error in the original measurement's approach like in \Cref{sec:dom:compariosn}) or \emph{exogenous differences in page re-execution}
(\EG{} browser non-determinism as discussed in \Cref{sec:back:limits}).

In summary, \tool{} is designed to be a tool for \emph{general}, \emph{accurate}, and \emph{comprehensive} \Web{} security
and privacy measurements, and produces \format{} archives designed to enable
accurate, reusable, general data archiving to aid reproducibility of \Web{} measurement.

% There rest of this section provides a brief description of \tool{}'s design, and the structure of \format{} archives.

\subsection{\tool{} Design}
\label{sec:tool:tool}
We implement \tool{} as a modified Chromium browser. \rev{As explained in \Cref{sec:back:tools:instrumentation}, this abstraction level is needed to ensure \emph{accurate}, high-fidelity insights into website executions.Chromium's market share of over 70\%~\cite{browserMarketShare}, makes it a very \emph{general} and long-term solution. Naturally, this decision comes with the limitation that \tool{} is browser-dependent. However, we do not focus on cross-browser differences but rather on capturing \emph{accurate} and \emph{comprehensive} \Web{} behavior. Alternatively, supporting all major browsers would add significant complexity and maintenance challenges.}

\tool{} consists of three parts: i. a system for determining \emph{which} page(s) is measured, ii.
a system for measuring \emph{what} resources are fetched during page execution, and iii.
a system for measuring \emph{how} the browser behaved while executing the page. Each capability is
briefly described below.

First, a researcher directs \tool{} to measure a webpage in one of two ways: either first, by using
the existing, well-known, and previously discussed Puppeteer library to pragmatically navigate \tool{} to a page to measure, or second, by
using \tool{} in an interactive session, where the researcher uses the browser as a user typically would,
but with \tool{} recording all relevant page behaviors.

Second, \tool{} measures what resources are fetched during page
execution using Chromium's built-in 
\emph{DevTools} protocol, the same
protocol used by Chromium's debugging tools and Puppeteer to
automate Chromium instances.

Third, \tool{} records browser behavior during page execution by instrumenting Blink,
the open-source system used for parsing HTML, presenting Web pages, and
handling the interaction between \JS{} and browser APIs used in Chromium.
Our approach builds on an existing Chromium-based measurement system called PageGraph, which has been used in
prior work~\cite{chen2021detecting,smith2022blocked,smith2021sugarcoat}.
PageGraph \rev{\emph{comprehensively}} tracks and attributes all network requests, DOM changes, and denoted
WebAPI and JS-builtin calls, building an internal graph of annotated \emph{actors}, 
\emph{actees}, and \emph{actions}. \tool{} extends PageGraph in a number of ways,
including 
% adding support for SVG sub-documents, FH: I removed SVG, because we later 4.4.2 we mention that event handlers in SVG were not detected at the time of our experiment.
handling and recording redirection in
HTTP requests and recording the arguments to (and structured responses from) WebAPIs, among other changes needed to support the high-fidelity and retrospective use cases \tool{} aims to target.

Finally, we note several attributes of \tool{}'s construction that better ensure accuracy
and help to avoid some of the correctness issues that affect some ad-hoc, or per-project,
measurement tools. For example, \tool{} is validated against the cross-browser
Web Platform Tests~\cite{webPlatformTests} to ensure stability and coverage (\IE{} that 
\tool{} can handle and record the enormous range of browser behaviors used by modern
websites). Additionally, \tool{} is kept current with every major Chromium release,
ensuring that it accurately records the Web as executed in a then-current browser \rev{(see \Cref{sec:dis:maintenance})}.

In short, \tool{} is
built to minimize (or avoid) the limitations of other browser-modification \Web{} measurement tools.

% Finally, we note the following to contrast \tool{} with the earlier mentioned browser modification approaches. \tool{} is built with industry-driven
% domain expertise of Chromium architecture, has been regularly updated
% and kept current with over a dozen Chromium versions, and is tested for correctness and robustness
% against hundreds of Web Platform Tests~\cite{webPlatformTests}. It is design to align with Chromium's architecture, to ease maintaining, updating, and ensuring \tool{}'s correctness as much as possible. 
% % For example, \tool{} attributes Web API calls by instrumenting Chromium's code-gen subsystem\footnote{The step in the Chromium build process where C++ code is dynamically generated to allow page \JS{} to access C++ defined ``backend'' methods.}, traces network requests and redirects via Chromium's ``DevTools'' interface\footnote{\url{https://developer.chrome.com/docs/devtools/network}}, and records DOM element creation and modification through Chromium's notification-driven ``CoreProbe'' system.
% In short, \tool{} is
% built to minimize (or fully avoid) the limitations of other browser-modification \Web{}
% measurement tools.

\subsection{\format{} Archives}
\label{sec:tool:format}
Finally, we discuss the format and contents of the \format{} files \tool{} produces when measuring a page's execution.

A \format{} archive consists of three files. First, each it contains a screenshot of the measured website at the end of the measurement period (\IE{} when \tool{} stops measuring page behaviors and begins serializing the behaviors it recorded into a \format{} archive). This screenshot is captured using the Puppeteer library, similar to other \Web{} measurement tools.

Second, each \format{} archive includes a copy of resources fetched during the page's execution.
This archive is also collected using Puppeteer and is encoded as a HAR
archive, again, in a manner similar to existing measurement tools.

Third, and most novel, each \format{} archive includes a description of all page behaviors measured during the page's execution.
\rev{Rather than using existing archiving formats like HAR or WARC, we adopted PageGraph's graph-based format to represent element dependencies and action flows as a directed multi-graph (as GraphML XML). Compared to list-like formats used in traditional archives, this approach offers a more intuitive representation of action dependencies and flows.}
This graph records all \emph{actor-action-actee}
behavior tuples described in the previous section, capturing all observed communication, DOM-modification, and script behaviors (\IE{} calls to 
significant Web APIs and \JS{} builtins) that occurred during the page's execution, detailed enough to allow the page's execution to be accurately replayed,
step by step.
% Demonstrative examples of the kinds of metadata included in the graph as node and edge attributes include the frame id of the execution context that a WebAPI was called in, or the name of an attribute modified on a DOM element, or the URL a request was being redirected to, among many others.

In conclusion, each of the three files included in each \format{} archive fulfills a different purpose to
enable accurate, replicable, historical \Web{} measurement: the screenshot file records the presentation of
the archived page, the HAR file records the resources loaded, and the GraphML file records
the iterative behavior, over the page's entire lifetime.

\section{Accuracy Improvements}
\label{sec:accuracy}

% In the beginning of this paper, we have seen that existing tools and archives used in Web measurement studies suffer from various threats to consistency. With the novel included description of all page behaviour as part of \tool, we believe to counter these threats and give a more accurate picture of what researchers saw in their original page visit. In this section, we answer the question \textit{what accuracy improvements do we get from adding \traces as part of an archiving format?}
%We claim that \tool{} is way more accurate in reproducing previous experiments than existing archives.
This section demonstrates how \tool{} can be used and the improvements it offers compared to common HTTP archives, \revise{e.g., enhancing accuracy in reproducing prior experiments.}

To do so, we revisit and reproduce three representative Web measurement experiments from previous works, creating WARC, HAR, and \format{} datasets for comparison. We focus on three commonly measured resources: First, we analyze the accuracy of recording \textit{JavaScript (JS) executions} by replicating experiments from Snyder~\etal~\cite{snyderBrowserFeatureUsage2016}. Then, we compare accuracy in recording \textit{Document Object Model (DOM)} content, a commonly measured resource, \EG{} to analyze third-party roadblocks through inline event handlers for Content Security Policies (CSP)~\cite{Steffens2021}. Lastly, we measure how the \textit{HTTP requests} differ and check how this helps the reproducibility of privacy research on third parties and trackers~\cite{englehardt2016online}. Before the experiments begin, we need a robust measurement pipeline.

\subsection{Dataset Construction}

\begin{figure*}
\centering
\includegraphics[width=\linewidth]{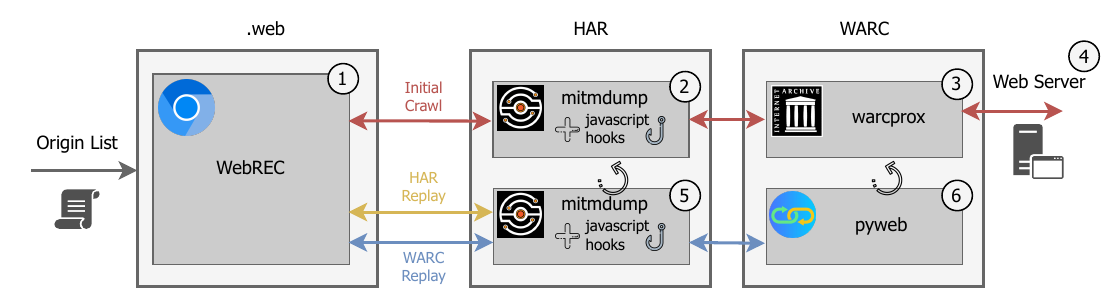}

\caption{The pipeline used to create our inital dataset plus datasets for HAR replays and WARC replays.
% \textcolor{purple}{The purple arrows path indicates the initial construction process}. \textcolor{blue}{Blue indicates the path for replayed WARC responses} and \textcolor{orange}{orange indicates the replay HAR responses path}.
}

\label{tikz:pipeline}
\end{figure*}

% In this section, we are interested in the comparison of different archive file formats against the PageGraph and the benefits of the PageGraph as an archive. 
To perform meaningful analysis and comparison between different archiving formats, we need a dataset construction pipeline that meets several criteria. First, we need a \emph{baseline} (BL) of recorded actions to compare all archives against. Otherwise, we cannot tell which archive is correct if two of them differ. Second, we require \emph{consistent} recording of the same events during a single visit \revise{to account for non-determinism~\cite{rothSecurityLotteryMeasuring2022}}. Third, the dataset needs to be \emph{representative} for common \Web{} security and privacy measurement studies. To meet these criteria, we built a pipeline that creates a BL dataset from requests sent to the sites from the CrUX top 10k bucket, \rev{a highly accurate and literature-recommended website popularity list~\cite{ruthWorldWideView2022}}. We proxy all requests through commonly used archiving tools to record the same requests (see~\Cref{tikz:pipeline} \revise{for the pipeline and \Cref{apx:ethics} for ethical considerations}).

Our pipeline begins with an origins list (CrUX, December 2023). For each origin, it initiates a new instance of warcprox~\cite{warcprox} (3), and mitmdump~\cite{mitmproxy} (2), and then runs \tool{} (1). We chose warcprox (from the Internet Archive) and mitmdump (part of mitmproxy) because they are well-maintained and tested tools. \tool{}'s requests are first proxied through mitmdump and from there through warcprox to the \Web{} server (ignoring TLS certificate errors). This setup ensures that all parts receive the same requests and responses, allowing us to record \format, HAR, and WARC files accurately.

\revise{With the \format, HAR, and WARC files, we can now compare requests and different file contents. However, it would not be possible to compare dynamic \JS{} executions on a page for HAR and WARC files. 
To measure executions on archived content, prior work has made use of sites like the Internet Archive that replay recorded responses~\cite{roth2020complex, stock2017web, lernerInternetJonesRaiders2016}.
%With a typical archive, one would usually measure executions by replaying the recorded server responses from the archive file.
We apply the same method by using the replay functions of mitmdump for HAR files (5) and pywb~\cite{pywb} (recommended by warcprox) for WARC files (6), and then measure executions by hooking into monitored \JS{} functions, keeping \tool{} as unmodified as possible. The hooks are set via a prepended script element in front of every HTML response with the mitm proxy (2 \& 5). Due to the error tolerance of browsers~\cite{hantkeHTMLViolationsWhere2022a}, this script is executed even though it is positioned in front of any HTML tag.} We note that this might lead to a difference in parsing due to the now entered quirks mode~\cite{quirksMode}, a point that we discuss in \Cref{sec:limitations}. Additionally, the script also hooks into programmatically created iframes. To ensure that the script is always executed and not blocked by CSP, we use the Puppeteer option \textit{setBypassCSP} to deactivate CSP. This approach maintains consistency by using the same tools for the initial crawl and the replays, differing only in the response source; archive files instead of the live web.
% , with the only difference being archive files as the response source instead of the live web.
% , our intuitive idea was to modify the pagegraph-crawl to inject JavaScript on each origin that records all requests and JavaScript executions inside the browser. However, this led to crashes of the browser. Instead, 

For our baseline dataset, we use the \JS{} (2) recorded with hooks during the crawl. Since \tool{} is designed to avoid altering browser behavior, the \JS{} hooks remain unaffected. After completing the pipeline for the CrUX top 10k, we have a solid dataset and baseline for analysis.

\subsection{Limitations}
\label{sec:limitations}

% \FH{TODO: merge this section into the flow}
Before focusing on the analysis, we first discuss the pipeline's limitations.
%Before we start with the analysis, we first must discuss our pipeline's limitations.
As previously mentioned, adding a script tag at the beginning of each document causes the browser to enter quirks mode~\cite{quirksMode}. The quirks mode is an approach by browser vendors to ensure backwards compatibility and render even severely malformed HTML with their \Web{} browser.
% to ensure backwards compatibility with Web pages that are older than modern standards.
It causes the browser to follow other parsing algorithms and leads to slight differences in the parsed results compared to what typical users would see (\EG{} CSS parsing). Yet, we are not aware of any relevant changes that would influence our results.

% Yet, to the best of our knowledge, this does not significantly influence our results as browsers are error-tolerant in parsing HTML anyway.

Furthermore, the pipeline misses executions inside of iframes added by the HTML \rev{parser} if they contain a \texttt{srcdoc} property. \rev{Using a custom query with the HTTP Archive~\cite{httparchiveBigQuery}, we can see that only 10,534 (0.9\%) of the 1,226,954 recorded iframes in their 1M dataset use \texttt{srcdoc}.} While our pipeline could detect and rewrite these during the rewriting phase with mitmproxy, it could cause side effects introduced by HTML parsing libraries (e.g., by removing malformed elements)~\cite{hantkeHTMLViolationsWhere2022a}. 
%some iframes we are unable to hook into. We cannot analyze iframes that are inside the main HTML document and use the srcdoc property instead of src, as we avoid parsing any content to avoid mutation-like side effects, which would likely impact our results far more.
Hence, we believe the impact of such mutations could be more significant to our results than being unable to hook \rev{into 0.9\% of the iframes} and thus decided against this approach.
%our findings should be considered a lower bound.

\revise{Finally, given the multi-tier architecture, there is a small chance for a race condition in the resource recording. This causes a slight mismatch in recorded resources between archives and the \format{} format, leading to a very small difference in reported resources (details in \Cref{sec:requests:correctness}).}
%Lastly, the pipeline has a race condition that occurs if server responses take longer than the time needed to generate the archive files.

% Besides limitations of our pipeline, there are also limitations in PG that have slightly influenced our results. First of all, PG only collects activities happening on the page, forcing us to filter out technologies like service workers that happen outside the context of a page. We describe in detail in the following sections how we attributed different contexts.
% Furthermore, PG is at the time of writing not able to records event handlers inside of inline SVGs, leading to few false negative reported cases with PG.

\subsection{JavaScript}
% \FH{TODO: Manually check the 163 origins for which PG has less executions than others. Likely only after submission as answer for rebuttal.}

%Now, we begin with the experiments.
Numerous Web security and privacy measurements~\cite{soMoreThingsChange2023, kleinHandSanitizersWild2022, snyderMostWebsitesDon2017} analyze \JS{} execution. In this section, we want to assess the differences, if any, in measured \JS{} executions between reproducing experiments using a traditional archival format and our proposed \format{} format. To assess this, we replicate experiments from Snyder~\etal~\cite{snyderBrowserFeatureUsage2016}, who investigated the prevalence if various \JS{} APIs. The authors injected \JS{} into each page to modify methods and properties to determine which API is used how often.

\subsubsection{\rev{JavaScript} Experiment}
Our experiment focuses on APIs related to two standards: \textit{CSS Object Model} (CSS-OM)~\cite{cssObjectModel} and \textit{HTML Channel Messaging} (H-CM)~\cite{htmlChannelMessaging}. Snyder~\etal report that CSS-OM is often used and rarely blocked by privacy tools, whereas H-CM is very actively used but frequently blocked. This small scope allows us to highlight key differences between archiving techniques.

%to two JS standards as examples to highlight differences between the archiving techniques: the \textit{CSS Object Model} (CSS-OM) standard~\cite{cssObjectModel} which is often used in the wild and rarely blocked by privacy tools according to Snyder, and the \textit{HTML Channel Messaging} (H-CM) standard~\cite{htmlChannelMessaging}, also a very actively used but frequently blocked technology.

% The functions of the these two standards are currently not recorded by the PG by default. To monitor them, we had to add them in the corresponding interface file
% \footnote{\url{https://sourcegraph.com/github.com/brave/brave-core@2796f71cfae057f4bfa9bb76d8d74dd1438f76d7/-/blob/chromium_src/third_party/blink/renderer/bindings/scripts/bind_gen/interface.py}}
% in the Brave codebase and build a custom Brave version to continue the experiment with.

To count the API invocations after running our pipeline, we first analyze \format{}'s execution graphs for \textit{js call} edges, which indicates API invocations. For the BL dataset and the replayed WARC and HAR responses, we count calls by analyzing the log files where the earlier mentioned \JS{} hooks log every \JS{} action on the page.

\subsubsection{\rev{JavaScript} Comparison}
After running the pipeline on the CrUX 10k bucket, our dataset contained 8,479 \format{} files, each indicating that a crawl was successful. Next, we replayed the responses from HAR and WARC files. For \empirical{272} and \empirical{75} origins, respectively, replaying failed due to timeouts in reading the file (although set to a generous 60 seconds) or file interpretation errors. 
%
%However, we also faced 272 and 75 issues, respectively, 
%due to timeouts in reading the file (although set to generous 60 seconds), or file interpretation errors.
Due to overlapping issues, this leaves us with 8,150 successfully crawled and replayed origins for the remainder of this section.

\begin{table}
    \centering    
    \footnotesize
    
    \begin{tabular}{lrrrrrr}
        \toprule
            Appearances & \multicolumn{2}{c}{\format{}} & \multicolumn{2}{c}{HAR} & \multicolumn{2}{c}{WARC} \\
        \midrule
        Equal &  39,603 & 93\% & 22,688 & 54\% & 22,645 & 54\% \\
        More & 2,918 & 7\% & 1,934 & 5\% & 1,725 & 4\% \\
        Fewer & 163 & 0.3\% & 17,557 & 37\% & 17,809 & 42\% \\
        \bottomrule
    \end{tabular}
    \caption{Appearances on the archived sites compared to 42,179 appearances on the baseline (BL).}
    \label{tab:javascript}
\end{table}

\begin{figure}
    \centering
    \includegraphics[width=\linewidth]{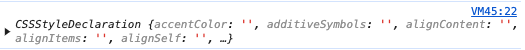}
    \caption{The DevTool logs show attributes of a CSSStyleDeclaration that are also logged with \tool.}
    \label{fig:devtools-log}
\end{figure}

For each origin, we counted the appearances of CSS-OM and H-CM API invocations. We define an \textit{appearance} on an origin as the combination of the API name and the number of times it was executed. Our BL dataset shows 15,775,881 API calls grouped to 42,179 appearances across origins. \Cref{tab:javascript} shows that out of these 42,179 appearances the \format{} dataset aligns with 39,098 (93\%) appearances on the same number of API calls, i.e., \tool{} recorded the same number of invocations as the BL. For 2,918 (7\%) of the appearances, \format{} captured more API calls, which can be attributed to \tool's deeper hooking capabilities compared to the hooks implemented in \JS{} capturing only surface-level activities. For instance, the code in \Cref{lst:devtool-hook-examples} shows a snippet in which the SVG's \textit{CSS Style Declaration} is logged, leading to the log output in \Cref{fig:devtools-log}. The lookup for attributes that appear in the log, e.g., \textit{accentColor} and \textit{additiveSymbols}, are also recorded in \format{} while the \JS{} hooks for the BL ignore it. Again, this difference is expected due to the underlying hooking differences. Moreover, only 163 (0.3\%) appearances had fewer invocations in \format{} than in the BL, showing that researchers can achieve a nearly identical representation of API invocations as in the original visit, even in replication experiments conducted years later.
% FH: I plan to look into these cases.
% Although the number is small, it is unexpected as the PG should record all executions. \red{Looking into these unexpected differences, it appears to occure very often for \textit{Window.getComputedStyle} for one single execution (the first execution?). In the logs, it seems that it could be an issue in our pipeline?}
Across all origins, the average difference in appearances between \format{} and BL is only 0.7\%, highlighting the approach's accuracy.
% \BS{How to compute the average here? }

After comparing the baseline to \format, we now focus on the comparison with replayed responses.
%Now that we have seen how \format{} compares to the BL, how do the replayed responses perform? 
Given that the underlying method to record the API calls is the same as for the BL, one would naturally expect results closer to the BL.
%a more accurate representations. 
%Contrary to this expectation, we assumed in our hypothesis a different outcome.
The data shows that only 22,688 (54\%) appearances in the HAR replays and 22,645 (54\%) for WARC align with the BL, i.e., the overlap is significantly lower than for \format{} (93\%).
On the one hand, both HAR and WARC also have cases in which more API invocations are detected compared to the BL.
We attribute this to the fact that replayed responses have no network delay when loading the resources, triggering some reoccurring API calls earlier.
% We attribute this to API calls, which happen at regular intervals, combined with the fact that no network delay in loading the resources occurs, due to the local replay.

\begin{listing}
    \begin{minted}[
        frame=lines,
        framesep=2mm,
        baselinestretch=1.2, 
        fontsize=\scriptsize,
        linenos,
        breaklines
    ]{html}
<svg id="s" width="100" height="100">
  <circle cx="50" cy="50" stroke="green" />
</svg>
<script>console.log(document.getElementById('s').style)</script>
    \end{minted}
    \caption{DevTools log a CSSStyleDeclaration.}
    \label{lst:devtool-hook-examples}
\end{listing}

\begin{figure}
    \centering
    \begin{subfigure}[b]{0.45\linewidth}
        \includegraphics[width=\linewidth]{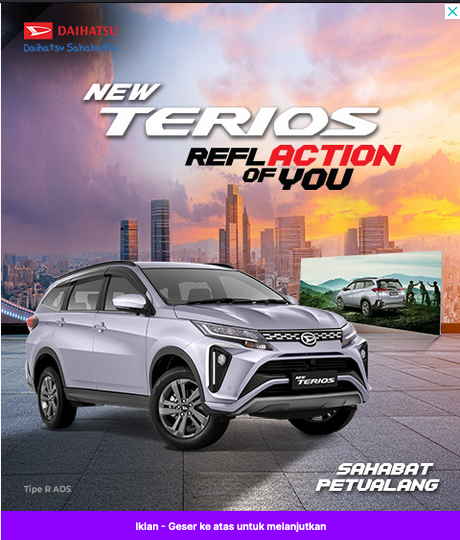}
        \caption{Ad from live website.}
        \label{fig:ad-live}
    \end{subfigure}
    \hfill % Optional: add some horizontal spacing
    \begin{subfigure}[b]{0.45\linewidth}
        \includegraphics[width=\linewidth]{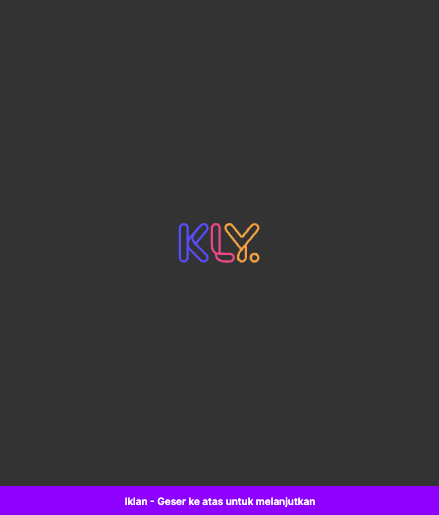}
        \caption{Ad from archived website.}
        \label{fig:ad-replay}
    \end{subfigure}
    \caption{Google ads load dynamic content leading to a difference of JS executions between live and archived websites.}
    \label{fig:ads}
\end{figure}

On the other hand, the vast majority of appearance differences are because \emph{fewer} invocations are detected in HAR and WARC; 17,557 (37\%) and 17,809 (42\%), respectively. 
Here, various of the reasons mentioned in~\Cref{sec:back:limits} could be at play. For instance, our dataset reveals that sites that use Cloudflare (\IE{} \emph{}{challenges.cloudflare.com}) for bot protection tend to have fewer API calls for the replayed responses, likely due to the lack of real-time interactions with their servers, which are crucial for dynamic content exchange and behavior measurement. Another example is shown in \Cref{fig:ads}, a Google ad element, first on the live visited website (\ref{fig:ad-live}) and then on the archived website (\ref{fig:ad-replay}). While the archived version shows only a static logo, the live version loads an advertisement after a few seconds from the ad servers leading to various CSS-related calls. Especially for privacy related research, this difference could be relevant. And this difference in appearances is significant: While we observed an average difference across all origins of 0.7\% between \format{} and BL, \emph{we measured 13.9\% for HAR and 13.3\% for WARC replayed responses}.
This highlights the drawback of replaying content rather than recording and inspecting the executions directly.

% \subsubsection{Replication}
% Now, we use our measurements to replicate the popularity measurement originally conducted by Snyder~\etal~\cite{snyderBrowserFeatureUsage2016}. While the authors analyzed 4 features for HTML Channel Messaging (H-CM) and 15 for CSS Object Model (CSS-OM), we found 5 and 57 in the according standards~\cite{cssObjectModel, htmlChannelMessaging}. At the same time, we only visited the landing page of each origin, while Snyder~\etal{} visited up to 13 pages per site providing a comprehensive dataset. \BS{Why are the 5 and 57 meaningful?}

% \BS{The absolute numbers are not explained, but also the percentages are not obvious to me. Also, why is the baseline lower than recorded? }
% Our findings (\Cref{tab:js-replication}) show consistency with previous work for CSS-OM with usage around 80\%, while H-CM popularity seems to decrease from around 50\% to less than 40\%. This trend is evident across all formats, indicating that, luckily, the bigger picture for studies replaying responses from HAR or WARC remains valid. Nevertheless, we recommend using \format{} for future work, knowing from the previous section that \format{} is more accurate in detailing page activity. Additionally, traditional formats require complex pipelines to measure JS and replicate experiments, while \format{} can easily be generated with \tool{} and analyzed in one run increasing efficiency and reducing the potential for errors (like the 347 timeout and interpretation errors).

% \input{tables/javascript_replication}

\subsection{Document Object Model}
\label{sec:dom}

Another typically analyzed component in measurement studies is the DOM of a webpage and how it behaves, for example, in case of DOM Clobbering~\cite{khodayariItDOMClobbering2023}, or with invalid HTML~\cite{hantkeHTMLViolationsWhere2022a}. To showcase how \format{} could be used for such experiments, we replicate a measurement from  Steffens~\etal~\cite{Steffens2021}, in which they analyzed to what extent third-party content on webpages complicates the security mechanisms Content Security Policy (CSP) and Subresource Integrity. 
In this section, we replicate one part of their hypothetical what-if experiment, in  which the authors make the assumption that first-party developers want to deploy a CSP without using \texttt{unsafe-inline} and \texttt{unsafe-eval}. Their findings revealed that the vast majority of sites (86\%) are unable to deploy CSP due to \JS{} code, which adds inline event handlers to the DOM requiring \texttt{unsafe-inline}; we, therefore, focus in this experiment on JavaScript-added inline event handlers.

% In this section, we focus on CSP and replicate one part of their hypothetical what-if experiment. In their experiment, the authors make the assumption that first-party developers want to deploy a CSP without using \texttt{unsafe-inline} and \texttt{unsafe-eval}. Their findings suggest that the vast majority of sites (86\%) are unable to deploy CSP because of \JS{} code, which adds inline event handlers; we, therefore, focus on these JavaScript-added inline events handlers, which require \texttt{unsafe-inline}.

\subsubsection{\rev{DOM} Experiment}
\begin{figure}
    \centering
    
    \includegraphics[width=\linewidth]{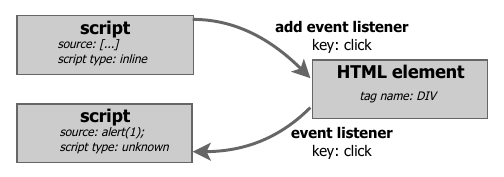}
    
    \iffalse
    \begin{tikzpicture}[node distance=0.1cm, auto]
    
        \tikzstyle{node} = [rectangle, minimum width=3.5cm, minimum height=1cm, text centered, align=center, draw=black, fill=blue!30]
        \tikzstyle{arrow} = [thick,->,>=stealth]
        \tikzstyle{labelstyle} = [text width=5cm, align=center]
        
        % Nodes
        \node[node] (script) {\textbf{script}\\\textit{...}};
        \node[node, below right= 1cm and -1.5cm of script] (element)  {\textbf{HTML Element}\\\textit{..}};
        \node[node, above right= 1cm and -1.5cm of element] (handler){\textbf{script}\\\textit{alert(1)}};
        
        % Edges
        % {'edge type': 'add event listener', 'id': '4916', 'timestamp': '8874', 'key': 'click', 'event listener id': 39, 'script id': 18}
        \draw[arrow] (script.south) |- (element.west) node[text width=3cm, midway] {add event listener;\\key: click};
        
        % {'edge type': 'event listener', 'id': '7153', 'timestamp': '20096', 'key': 'click', 'event listener id': 39})
        \draw[arrow] (element.east) -| (handler.south) node[text width=3cm] {event listener;\\key: click};
    \end{tikzpicture}
    \fi
    
\caption{One example how an EventListener is represented as script node in the execution graph.}
\label{tikz:event-listener-example}
\end{figure}

To analyze inline event handlers, we run Steffens~\etal{}'s open-sourced SMURF code in our pipeline to count all origins that use inline event handlers added to DOM through APIs like \texttt{document.write} and \texttt{innerHTML}.
Using this code, we can simply count the origins in our log files and use it as a baseline for the comparison. We use the same method for the replayed responses from HAR and WARC files.
Within \format, we query the graph for \textit{event listener} edges (see \Cref{tikz:event-listener-example}). If these point to the same JavaScript node that added the event listener (\IE{} \textit{add event listener} edge) or the script node does not have the type \emph{unknown}, they are \emph{programatically added}, i.e., do not involve string-to-code transformation. If, instead, they were added from the HTML parser or another script, they are actual inline event handlers, which would require \texttt{unsafe-inline}.

%In PG, we look for \textit{event listener} edges which point to the script the event listener executes and use \textit{add event listener} edges to understand who added the listener. One such example pattern is shown in~\Cref{tikz:event-listener-example}. However, we need to be careful not to add all event listeners to our count, as the same edges are also used for programatically added event listeners. For instance, if the \textit{event listener} edge is self referencing the same script node, it was programatically added via the \texttt{addEventListener} function which does not violate CSP. Similarly, if the script type of the event listener node is not \textit{unknown}, it also indicates a programmatically added event listener. Using these metrics, we can accurately distinguished between inline or programmatically added event listeners in PG.

%\BS{This is rather specific about pagegraph. Do we need the level of detail here or does it make more sense to rather focus on the issue with Steffens' code?}

\subsubsection{\rev{DOM} Comparison}
\label{sec:dom:compariosn}

In this crawl, we successfully visited \empirical{8,523} origins with \empirical{311} HAR and \empirical{44} WARC (overlapping) issues replaying responses, similar to the previous experiment. Thus, this sections builds
on
% \empirical{8,169}
\rev{8,170}
visits to analysis. 

The results of the analysis are shown in~\Cref{tab:unsafe}. At first sight, the results appear extremely surprising: the original version of \rev{Steffens~\etal{}'s code, SMURF~\cite{smurfCode}}, finds \empirical{four times} as many sites with inline event handlers
% (\empirical{6,190})
(\rev{6,188})
compared to \tool{} 
% (\empirical{1,545}).
(\rev{1,515}).
We, therefore, carefully analyzed SMURF's source code and found a number of impactful bugs. First, their code hooks into APIs which can be used to add elements to the DOM and subsequently checks if the added element contains inline event handlers. However, their code iterates over all the \emph{properties} of an element instead of checking for \emph{attributes}. This difference is critical, though: if an event handler is programmatically added through a function reference, this is a \emph{property} of the element but not an \emph{attribute}. In contrast, adding an inline event handler through \texttt{ele.setAttribute('onerror', '...')} (which requires string-to-code transformation) sets the attribute. A programmatically added event listener for an HTML node is a common way for many third parties to react to events while still allowing CSP to block inline scripts (a real-world example can be found in~\Cref{apx:google-analytics}). The SMURF code also does not properly capture event handlers which are nested inside a subtree of the DOM, leading to false negatives. Finally, SMURF assumes that any attribute starting with \emph{on} is an event handler; as a result, our initial experiments showed that due to typos (e.g., \emph{onclik}), 
HTML elements without meaningful event handlers were flagged as CSP blocking.
%HTML elements are incorrectly reported as having event handlers which would be blocked by CSP. 
Upon discovering of these three issues, we \rev{contacted the original authors and fixed the code together with them} to have a more meaningful baseline.

\begin{table}
    \footnotesize
    \centering    

    \begin{tabular}{lrrrrr}
        \toprule
        Method &  SMURF (bug) & SMURF (fix) & \format{} & WARC & HAR\\
        \midrule
        % old: Origins &  \empirical{X,XXX} & 1,725 & 1,549 & 1,441 & 1,552 \\
        % Origins &  \empirical{6,190} & 1,645 & 1,545 & 1,473 & 1,368 \\
        Origins &  \rev{6,188} & \rev{1,644} & \rev{1,515} & \rev{1,472} & \rev{1,368} \\
        \bottomrule
    \end{tabular}
    \caption{Origins for which we found inline event handlers.}
    \label{tab:unsafe}
\end{table}

Given the updated code, we now investigate the feasibility of WARC, HAR, and \format{} to a study such as Steffens~\etal. We find that the replayed responses from WARC and HAR show fewer origins with inline event handlers, 
% \empirical{1,473} 
\rev{1,472}
and 
% \empirical{1,368}
\rev{1,368}
respectively, than the initial baseline of 
% \empirical{1,645}
\rev{1,644}
origins. This discrepancy is again due to dynamic elements, such as \JS{}, not reloading accurately.  These findings reinforce our argument that archive-with-metadata approaches are not ideal when the goal is to accurately reproduce Web measurements.

For \tool, there remains a difference from the baseline
% (\empirical{1,645} vs. \empirical{1,545}).
(\rev{1,644} vs. \rev{1,515}).
Closer analysis of the findings showed that one issue relates to PageGraph, which was not able to handle event handlers inside SVG elements. \revise{We reported this limitation to the authors who addressed it in a newer version}. Moreover, SMURF just logs any event handler regardless of whether it was added to the page's DOM and triggered (media elements are an exception) and irrespective of their origin, i.e., also reports those added in cross-origin iframes, which are not relevant for CSP. Hence, we believe the fixed implementation of SMURF to be in line with what \tool{} can record.

\subsubsection{Consequences}
In order to conclusively refute the findings of Steffens~\etal, we would have to apply the fixed version of their code to the dataset from 2020. However, without a published dataset that is accurately reproducible like \tool, we cannot make ultimate statements; even if the authors had recorded HAR or WARC files, it would be infeasible to fully replicate their findings.
Nevertheless, we believe that inline event handlers played a less significant role for CSP blockage (assuming the Web has not radically changed since 2020). However, the authors also crawled sub-pages that could contain more event handlers and analyzed other aspects like inline scripts or eval calls, which we did not investigate. This makes us believe that the general takeaway of the paper remains valid.

%Since replicating the experiment by Steffens~\etal{} is now obsolete, we rather take a moment to consider what the consequences of our finding are. For the conclusion of the original paper, we assume that inline event handlers play a less important role in blocking the usage of CSP than we thought so far. However, without a published dataset which is accurately reproducible like \tool, we cannot make ultimate statements. 
%Furthermore, the paper also crawled sub-pages that could contain more inline-event handlers and analyzed other aspects like inline scripts or eval calls which we did not investigate. This makes us believe that the general takeaway of the paper remains valid.

In general, our findings show that the complexity of Web measurements and JavaScript hooking is prone to cause errors.
%For Web measurement in general our finding shows how complicated and prone to errors such experiments are due to the complexity of JS. 
Additionally, mistakenly collecting all \textit{on*} attributes or missing newly introduced event listeners cause problems in such traditional \Web{} experiments. As \tool{} records all activities on a webpage, the list of event listeners is always accurate. Consequently, we believe researchers utilizing this technology would significantly improve their experiment pipeline by reducing their chance of implementing bugs.
% Consequently, we believe researchers utilizing this technology would significantly reduce their chance of implementing bugs in their experiments.

% \input{figures/dependency-chain-example}

\subsection{Requested Resources}
\label{sec:requests}
% \BS{It's not really requests if we talk about security headers, but rather the response to that. Maybe say "Requested Resources"? Also, the section needs some more structure as we currently have a lot of text which covers both the filtering/"correctness" part as well as the part about replication}

Requests are frequently analyzed in Web security and privacy studies for various reasons, for example, to support arguments about security headers~\cite{calzavara2020tale}, to identify privacy leakages~\cite{senol2022leaky}, or to measure online advertisement and tracking behavior~\cite{englehardt2016online, lernerInternetJonesRaiders2016}. For this section, we take the online tracking paper by Englehardt and Narayanan \cite{englehardt2016online}, a study in which the authors captured requests from one million websites and utilized the two tracking-protection lists Easylist and EasyPrivacy~\cite{easylist} to measure the prevalence of third-party requests and online trackers. Following their method, we can assess how well we can perform such an experiment using \tool{} and show what additional benefits researchers could get from it.

\subsubsection{\rev{Resources} Experiment}
% Numbers from hwpg/analysis/scripts/requests_results_1714863015_blocklists.csv"

Unlike the experiments before, researchers would typically not replay responses since all requests are stored in the HAR or WARC files already. Hence, we follow this method and only compare \format{} against HAR in this section without replaying responses. Additionally, we want to demonstrate how comprehensive \tool{}'s recorded page behavior is and use only this feature from \format{} to compare against HAR.

\begin{figure}
    \centering
    
    \includegraphics[width=\linewidth]{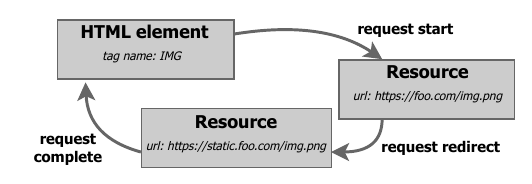}
    
    \iffalse
    \begin{tikzpicture}[node distance=2cm, auto]
    
        \tikzstyle{node} = [rectangle, minimum width=3cm, minimum height=1cm, text centered, align=center, draw=black, fill=blue!30]
        \tikzstyle{arrow} = [thick,->,>=stealth]
        \tikzstyle{labelstyle} = [text width=5cm, align=center]
        % Nodes
        \node[node] (element) {\textbf{HTML Element}\\\textit{IMG}};
        \node[node, minimum width=5.5cm, above right=of element] (resource1) at (0,0) {\textbf{Resource}\\\textit{https://foo.com/img.png}};
        \node[node, minimum width=5.5cm, below right=of element] (resource2) at (0,0) {\textbf{Resource}\\\textit{https://static.foo.com/img.png}};
        
        % Edges
        \draw[arrow] (element.north) |- (resource1.west) node[text width=3cm, midway, above] {request start};
        \draw[arrow] (resource1.south) -- (resource2.north) node[text width=3cm, midway] {request redirect};
        \draw[arrow] (resource2.west) -| (element.south) node[text width=3cm,midway, below] {request complete};

        % \node[labelstyle] at ($([xshift=2cm]script.east)!0.5!(element.north) - (-1.2cm, -0.5cm)$) {js call\\args: content-type};
        % \node[labelstyle] at ($(element.west)!0.5!(script.north) - (-1.2cm, +0.5cm)$) {js result\\value: text/plain;};
    \end{tikzpicture}
    \fi
\caption{Resource requests and redirects in the execution graph are represented as a resource nodes with edges.}
\label{tikz:redirect-example}
\end{figure}

In these behavior recordings, we have two types of requests. Requests directed to load documents are tracked as URLs in the document node, i.e., the HTML pages themselves or any embedded document. Besides document requests, \tool{} records resource requests and represents them via an edge to a resource node as demonstrated in \Cref{tikz:redirect-example}. 

% Our research revealed a bug within PG that affected the tracking of resources whenever the requests were redirected. Upon disclosing it to the vendor~\cite{redirectIssue}, they addressed the issue so that now redirects are accurately recorded using a dedicated edge in the graph as also shown in \Cref{tikz:redirect-example}. 

\subsubsection{Correctness}
\label{sec:requests:correctness}

Our results show that the requests in \format{}'s behavior records have great agreement with those recorded in HAR. In total, we count \empirical{778,500} requests in HAR after filtering out non-page-related requests like service workers or initial redirects, while we see \empirical{776,229} requests captured via \tool{} for \empirical{8,544} successfully crawled origins. This makes a negligible difference of less than \empirical{0.3\%}. 
% \BS{You start with talking about filtered requests, but explain only later that things are different. Can we rather start with the absolute difference (which is much bigger) and then explain classification of reasons; this way, we can explain the limitations of PageGraph and then arrive at the conclusion that on 0.3\% are different.}
% \FH{We decided to keep it in the current flow and make clear in the first part that we filter out cases to get 0.3\%.}

% \BS{Not sure that "fair" is the right term. I would rather say: PageGraph covers requests from the page. These do not include things like Service Workers, pre-flight requests, CSP reports. Also, initial redirects are out of scope (as they do not belong to the page). Then explain that we can attribute many types of requests (e.g., Sec-Fetch-...). For SW, this doesn't work, which is a limitation of PG. To nevertheless attribute SW requests, we simply filter them out. }
% Despite the low number of differences, we still see a difference and also mention that we need to filter requests.
Filtering the HAR file is required due to the very different ways these two file formats work. HAR is developed to record \textit{all} requests that leave the browser, while \tool{} captures every activity happening \textit{on the page}. These do not include activities like service workers, pre-flight requests, WebSocket upgrades, or CSP reports. Also, initial redirects are out of scope as they do not belong to the \emph{page}.

We can attribute most requests made by the browser and filter them out from the HAR dataset by relying primarily on request headers. We attributed \empirical{12,342} pre-flight requests looking for the \texttt{Access-Control-Request-Method} header and \empirical{1,210} CSP report requests using \texttt{Sec-Fetch-Dest} headers set to \texttt{report}. \tool{} behaviors do also not record the upgrade request for a WebSocket which we attribute via the \texttt{Upgrade: websocket} header, leading to \empirical{1,455} requests being filtered out. To attribute initial redirects from a document, e.g., \textit{mobile.example.com} to \textit{example.com}, we used the Sec-Fetch-Dest \texttt{document} and the \texttt{Location} response headers which gave us \empirical{2,819} filtered requests. Additionally, Chromium itself causes more than \empirical{40} requests per session, e.g., requests to the updater and similar, leading to \empirical{449,511} requests, which we attribute by looking for these URLs. Lastly, service workers also live in their own context and are not recorded in \format{} behaviors.
While the initial loading of a service worker is detectable through the Sec-Fetch-Dest header, scripts included by the service worker do not carry such distinctive features. Therefore, we instead hooked into the serviceWorker API to disallow registration of any service workers.
%Based on the request or response headers alone, we could not attribute requests made by a service worker, thus we simply deactivated them in our experiment by hooking and neutralizing the WebAPI \texttt{navigator.serviceWorker.register}. 
% This led to the blocking of \empirical{1,325} service workers. 
% \BS{If we can get the numbers of contacted hosts, we explain this here, otherwise drop the sentence}

Besides requests outside the page context, we also see requests that are internally made by the page but that never leave the browser. These never touch the HAR proxy but are recorded via \tool.
For example, some \texttt{Non-Authoritative Information} responses, like HSTS upgrades, are internally represented as \texttt{307 Internal Redirect} or the WebRequestAPI, likely Chromium internal behavior. Lastly, our manual investigation into differing requests shows a small race condition, a limitation in our pipeline: 
Sometimes, the browser sends a request just before we generate the \format{} output and tear down our pipeline. This means, a response might not return in time, resulting in an unrecorded HAR entry but an existing request in the behavior recording. Alternatively, the request might return and be recorded in HAR, but only after the \format{} is generated, leading to one more request in HAR.
% If the browser sends a request just before we generate the PG, the request is already recorded in the PG as an outgoing request. Two outcomes are possible, either the response returns before we tier down our pipeline, then we have only observe an outgoing request in PG but also a response in the HAR file. Or, if the request does not return, the outgoing request is recorded in the PG but not in HAR, as the format only records finished requests. In our analysis, we decided to only count requests that were finished before the PG was generated leading to some more racy requests in HAR.
% \BS{The race condition is a bit convoluted: make it concrete: browser makes requests right when we generate PG and tear down. This means the request might not even go out to be HAR-captured (one more requests in PG than HAR) or comes back to HAR but not browser (PG has one less response in PG)}
Due to these racy requests and the page-internal ones we could not to filter out, we measure a difference of less than \empirical{0.3\%} in the total number of requests.

% From this requests comparison between PG and HAR, we can see that both formats are great for what they aim for; HAR is great for recording \textit{all} requests that leave the browsers, while PG is great for everything that happens \textit{on the page}. Depending on the experiment one plans to conduct, one should carefully choose the right format. For most web measurements, we assume it is essential to analyze what happens on the page. This is also the case with the following third-party and tracking experiment which shows very similar numbers for both formats.
% \BS{Should we really say that HAR is great? Modulo the fact that you have raw data, the space requirement is huge. If we want to "sell" PG, we should rather state that PG is fine for most types of measurements except for edge cases with SW.}

% \rev{Need rephrasing: } From this comparison, we can see that \tool{} is well-suited for research measuring activities occurring directly \textit{on the page}. At the same time, HAR generated via a proxy stores \textit{all} requests, resulting in a substantial space consumption; on average, a HAR file in our dataset takes around 40MB, compared to \format{}'s 12MB. We assume that most \Web{} measurement scenarios do not require the exhaustive list of all requests, which only consumes resources. In the next sections, we emphasize this by replicating the third-party and tracking experiment, showing very similar numbers for both formats, although \format{} stores only page context requests.

\rev{This comparison shows that \tool{} is well-suited for measuring activities directly \textit{on the page}, while the proxy-generated HAR captures \textit{all} requests -- information, often not needed for many studies, as the next section shows.}

\subsubsection{Replication}
\rev{This section replicates a common third-party experiment to demonstrate that the page-context-only approach produces nearly the same outcomes as the more inflated HAR.}

In the more than 700,000 requests we made to the top 10K websites, we see that \empirical{5,938} third parties are present on at least two first parties in the \format{} behavior records. At the same time, in HAR recordings, after filtering out the document redirects to avoid false positives (i.e., the false origin), we see \empirical{6,043} third-party requests.
% (which we need filter out as otherwise we would generate false positive results calculated in a false context, i.e., the false origin). 
Same as for previous work, we can see that the majority of third parties are only present on a few sites as only \empirical{719} (\format) and \empirical{739} (HAR) third parties are present on more than 1\% of all sites in our dataset. % (HAR with redirect filter)
% The numbers demonstrate that although we use very different formats to measure them, the results for PG and HAR are very much alike.
% , or \empirical{5,929} with the additional CSP report, websocket and preflight filters applied.
% \empirical{717} (HAR filtered),
Next, we measured the prevalence of the third parties and compared the results to tracking-protection lists. The results for the top 20 third parties are demonstrated in \Cref{{fig:third-parties}}, showing the same observations as made by Englehardt and Narayanan \cite{englehardt2016online}. As explained in their paper, the top third parties belong to only a small group of organizations. Indeed, the top 8 alone are associated with Google. Additionally, not all third parties are used in a tracking context, such as content delivery websites like \textit{gstatic.com} or \textit{jsdeliver.net}. The data confirms that whether measuring requests from the HAR recordings or utilizing activities captured with \tool, the overall picture and takeaways remain consistent since the original study.

\subsubsection{Attribution}
In some experiments, simply identifying the request-target might not be enough.
% For example, if a sandboxed iframe causes a request to a tracker, it has less information about the logged-in user.
\revise{For example, if we want to understand the supply chain of third-party dependencies, we need to properly attribute the initiator of a request.}
However, it is non-trivial to accurately attribute requests to the corresponding source from which they were requested if the measurement is only based on recorded requests. Of course, indicators like the Referer header offer some insights, yet the Referrer-Policy could skew these signals. Consequently, a thorough understanding of third-party requests necessitates insights into on-page activities, activities recorded with \tool{}.
%\BS{I have a hard time buying the argument about sandboxed iframes. We might need a better example. }
%The numbers from Engelhardt's experiment could change when considering the context from which a request was sent. Depending on the research question one tries to answer, the context could influence the results. For example, one might need to categorize a third-party request sent by a sandboxed iframe differently than a third-party request from the initial website. 
%Yet, it is impossible to accurately attribute requests to the corresponding first party from which they were requested if the measurement is only based on recorded requests. Of course, indicators like the Referer header offer some insights, yet the Referrer-Policy could skew these signals. Consequently, a thorough understanding of third-party requests necessitates insights into on-page activities, activities that are recorded with activity traces like \tool.

%To showcase the flexibility of \tool{} for such an analysis, we 

%In our experiment, we attribute the two most prominent third parties to their contexts, to illustrate the process of request attribution. We leave a more comprehensive analysis for future research. Nevertheless, this is enough to demonstrate what additional request insights the PG can give over traditional HTTP archives.
%\BS{How does this connect to the rest?}

\begin{figure}
    \centering
    \includegraphics[width=\linewidth]{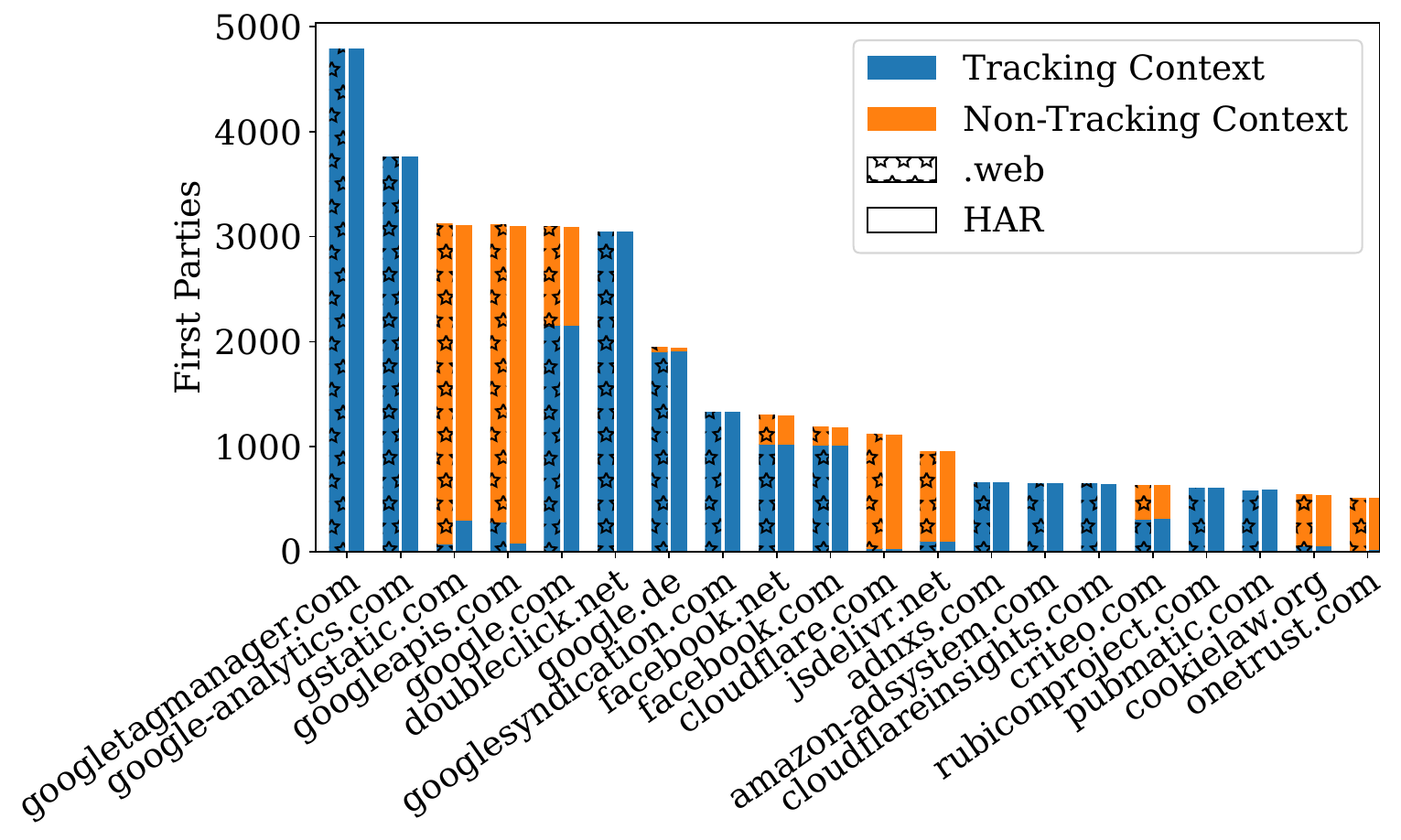}
    \caption{Top third parties on the top 10K sites via HAR and \format{}, divided in tracking and non-tracking context.}
    \label{fig:third-parties}
\end{figure}

%For the final investigation, we took the PG of each origin for which we earlier found requests to either \textit{googletagmanager.com} or \textit{google-analytics.com}. Google Tag Manager (GTM) is a tool that allows website operators to manage and deploy various JS modules and snippets, while Google Analytics is a web analytics service. For both, we queried the PG files for these requests and traced all edges to the DOM root node to generate a dependency chain. In total, we found \empirical{12,564} requests to the GTM and \empirical{11,298} to Analytics.

We now demonstrate the additional insights that \format{} can provide compared to traditional HTTP archives. To that end, we analyze the requests sent towards the most prevalent third party, i.e., Google Tag Manager (GTM), which received a total of \empirical{12,564} requests from \empirical{4,642} origins.

\Cref{tab:third-party-details} shows different variants through which the GTM is requested. Unsurprisingly, most requests come from script elements adding GTM. The most responsible party (causing \empirical{5,130} requests) is GTM itself, as the initial GTM script often includes additional scripts via script elements. It is followed by the script elements added by the first party, the expected way to add GTM in the first place.
The long tail of third parties responsible for script elements (\empirical{736} different third parties; not shown in the table), demonstrates the wide variety of entities websites use to load \JS{} modules, including alternative tag managers like Adobe Dynamic Tag Management or own content delivery network such as \textit{xhcdn.com}.

\revise{GTM is also used to load images (e.g., tracking pixels), mostly requested by GTM itself (563). Requests made through JavaScript fetch or XMLHttpRequests APIs towards GTM are mainly caused by the first party (157). In addition, 92 requests were caused by parallel-loaded script elements (\EG{} \emph{defer}) within the page's original HTML (\emph{Parser}).}

This shows that \tool{}'s behavior recordings can be used for both to accurately measure requests and to further understand why and how these requests occurred by providing far more details than a basic request record does, which would have to rely on heuristics to determine this metadata.

\begin{table}
    \centering       
    \footnotesize
    
    \begin{tabular}{llrr}
        \toprule
        Context & Responsible Party &\# Req. &\# Origins \\
        \midrule
        Script Element & \textit{googletagmanager.com} & 5,130 & 2,751 \\
         & \textit{first-party} & 4,342 & 3,503 \\
         & \textit{google-analytics.com} & 593 & 445 \\
         & \textit{adobedtm.com} & 77 & 40 \\
         & \textit{xhcdn.com} & 60 & 60 \\
         & \textit{...} & 1,534 & 1,169 \\
        \midrule
        IMG Element & \textit{googletagmanager.com} &  563 & 96 \\
         & \textit{halodoc.com} &  8 & 1\\
         & \textit{googleoptimize.com} &  1 & 1\\
        \midrule
        Fetch/XHR & \textit{first-party} & 157 & 1151\\
         & \textit{sports.ru} &  2 & 2\\
         & \textit{thesimsresource.com} &  2 & 1\\
         & \textit{geekdo-static.om} &  1 & 1\\
        \midrule
        Parser & \textit{first-party} & 92 & 74 \\
        % \midrule
        % This was an internal redirect (HSTS). Not interesitng for the reader.
        % Resource & \textit{googletagmanager.com} & 2 & 2 \\
        \bottomrule
    \end{tabular}
    \caption{Third party requests to GTM grouped by  context and responsible party who sent the request.}
    \label{tab:third-party-details}
\end{table}

\section{Generality of \tool}
\label{sec:availability}

Previous sections showed that common Web measurement methods lack reproducibility and sometimes correctness. We proposed \tool{} as a general-purpose tool to overcome these issues, \revise{yet its generalizability remains in question}.
% Nevertheless, the question remains, how generalizable \tool{} is for common Web studies.

To evaluate generality, we surveyed previous \Web{} measurement studies from top security and privacy conferences, building on a SoK paper by Stafeev and Pellegrino~\cite{stafeevSoKStateKrawlers2024}, who analyzed the use of crawlers in \Web{} measurement studies. \revise{As detailed in \Cref{apx:papers} we consider a subset of their reviewed papers, including studies from 2020 to 2022 that measured data within the scope of what WebREC offers, \IE{} they either analyze webpage JavaScript executions, the page’s content, or HTTP requests made. Thus, we focus on a set of \empirical{97} papers.}

\revise{Our survey shows encouraging numbers, suggesting that \empirical{68} (\empirical{70\%}) of the \empirical{97} papers could use \tool{} for most of their data collection. This would not only standardize the process and reduce the chance for errors during collection but also save researchers resources and time. For example, Musch and Johns~\cite{musch2021u} investigated \JS{} anti-debugging techniques in the wild by identifying code patterns and assessing performance changes, a tremendous development effort. Since these techniques are registered during page load, \format{} contains them, allowing this analysis without a custom crawler.}

\revise{On the other side, studies unsuitable for \tool{} include those measuring something outside the page context, \EG{} service workers~\cite{karami2021awakening} or browser extensions and those focusing on very specific features, such as DNS over QUIC~\cite{kosek2022dns}. Additionally, since \tool{} is based on Chromium, studies requiring other engines like Firefox are not catered for~\cite{narayan2020retrofitting}.}

Even though \empirical{70\%} of papers could rely on \tool{} instead of a custom crawler,
%While \empirical{70\%} is already a good number of papers that would not need to use a custom crawler,
the researchers would still require crawling, leading to avoidable resource usage and comparability issues. 
% Hantke~\etal~\cite{hantke2023you} suggested to base studies on web archives instead. 
In fact, our investigation also shows that \empirical{47} (\empirical{48\%}) papers could base their analysis solely on a \format{} archive instead, without the need for them to run a crawler.
For example, Luo~\etal~\cite{luo2022scriptchecker} conducted an experiment labeling DOM elements as sensitive, subsequently checking what third-party scripts access them. Calzavara~\etal~\cite{calzavara2021reining} studied inconsistencies in security mechanisms comparing cookie policies and security headers across sites. Both DOMs and headers are collected by \tool{} and could be analyzed in a \format{} archive.
 
To conclude, we believe that a significant percentage of Web measurement studies (\empirical{70\%}) could benefit from using \tool{} over custom crawlers and that \empirical{48\%} could even be based solely on a \format{} archive. This highlights \tool{}'s potential to streamline measurement processes and to enhance reproducibility and correctness in our academic field.
\section{Discussion}

With the results of the previous section in mind, we now summarize the primary benefits of \tool. 

%  HH: removing this as we mention these already 
%As outlined in \Cref{sec:back}, there has been an extensive development in tools and archive for web measurement, evolving from simple HTTP requests up to modified browsers. Despite these advancements, existing tools still face limitations, especially in terms of reproducibility, correctness and efficiency. With \tool, we aim to solve these challenges.

\subsection{Reproducibility}

\tool{} addresses the problem of reproducibility by providing a comprehensive bundle of information about page requests and activities for researchers to use and share. Recorded at the core of the browser, this information offers a much more accurate representation of page behavior than existing archiving formats. For instance, for \JS{} API calls, \tool{} shows an average difference of only \empirical{0.7\%} from the baseline, in contrast to a \empirical{13\%} difference with common HTTP archives. This discrepancy arises from dynamic behaviors like live server interactions, which can never be captured accurately by simply replaying responses.
\tool{} eliminates this need for replaying responses by storing all relevant activity information at runtime. Researchers can reproduce experiments by running the same scripts used in the original study on their dataset, ensuring reliable replication.
% Additionally, the approach also avoids encoding or timeout errors with parsing archives (which we saw in our experiments), again enhancing reproducibility.

Going one step further than reproducing experiments, \format{} also allows researchers to delve deeper into the initially recorded data, offering a better understanding of various aspects of a historical experiment. We demonstrated for the requested resources experiment that we can not only replicate the key findings of the original paper but we also easily attributed third-party calls to different scripts, providing valuable insights into third-party deployment. This advanced capability can help researchers explain why replicated findings might differ over time, e.g., due to changes in the mixed content policy. We encourage researchers to provide their measured data as \format{} archives for future research to use.

\subsection{Correctness}

Since \tool{} records behavior at the browser's core, the recorded behavior is correct and eliminates the need for researchers to modify browsers or inject code into \Web{} pages.

Custom modifications can introduce issues if researchers are not fully aware of every browser-specific detail, leading to unintended consequences. For example, 
%as discussed in \Cref{sec:dom},
we showed, how a study~\cite{Steffens2021} mistakenly attributed programmatically added event handlers as inline ones due to a mix-up between \emph{properties} and \emph{attributes}. \tool{} avoids such problems by recording all page activities accurately at its core, allowing researchers to analyze \format{} without an in-depth understanding of the HTML or other specifications. In \format{}'s execution graph, the source of an event handler is clearly visible, reducing the risk of misattribution during analysis significantly.

Even if a potential bug occurs in the analysis phase, other researchers could easily reconstruct and correct it based on the provided \format{} bundles.
% For Steffens~\etal~\cite{Steffens2021}, reproducing their experiment with our corrected script is not possible as their dataset is not available.
That this is desirable shows the Steffens~\etal~\cite{Steffens2021} paper, for which we could not rerun their experiment with our corrected script due to the lack of an available dataset.
Thus, we only run their script on our smaller dataset years later showing that over \empirical{50\%} of the visited sites in our data would have incorrectly attributed event handlers. How this would change the takeaway message of the original paper is unclear. We believe providing a format that allows other researchers to rerun and even correct experiments significantly improves the correctness of Web measurement research.

\subsection{Efficiency}

\tool{} improves the efficiency of Web measurement research. The experiments in \Cref{sec:accuracy} demonstrate how \format{} supports various studies, eliminating the need for resource-intensive development of custom crawlers. Around 70\% of papers we reviewed could use \tool{}, highlighting the benefits of a standardized crawler for our community. On top of that,  48\% of these papers conduct experiments that could rely solely on \format{} archives without the need to live-crawl. To make this possible for the community, we plan a publicly available, actively maintained \format{} archive for future researchers (see \Cref{apx:availability}). We are confident that this will save resources, reduce \Web{} traffic, and prevent potential ethical pitfalls.

Such an archive would allow extensive future work, like longitudinal \JS{} behavior analysis, currently not feasible with traditional \Web{} archives. It also allows easy prototyping of experiments without the need for additional crawls, thereby improving efficiency in \Web{} measurement research.

\subsection{Storage Implications}

\rev{Storing more information is expected to result in larger file sizes, requiring researchers to balance the new insights gained with storage demands. To address this, we added an optional compression step to our crawler and query tool. To estimate the expected storage requirements, we conducted another experiment comparing \format{} with the HAR files generated by a proxy. We ran our pipeline without any \JS{} injections (which would otherwise inflate the required space) on the CrUX 10k list, ending with 8,934 origins to compare.

For these origins, the average HAR file size was 13.6 MB, while \format{} files averaged 10.2 MB, broken down as 8.2 MB for HAR files, 1.5 MB for behavior records, and 0.4 MB for screenshots. The difference in HAR and \format{} HAR sizes is due to the fact that proxy-created HAR files include all browser requests, such as upgrade requests or updating dictionary files, whereas the \format{} HAR file only captures requests originating from the page context. This shows that \tool{}, while providing more insights, keeps storage needs low.
% For these origins, the average HAR file size was 13.6 MB, while \format{} files averaged 19.6 MB, broken down as follows: 8.2 MB for HAR, 1.5 MB for behavior records, and 0.4 MB for screenshot. The difference in the original and \format{} HAR file sizes is due to the fact that the HAR files created via the proxy include all browser requests, such as upgrade requests or updating dictionary files, whereas the \format{} HAR file only capture requests originating from the page context.
% If we compress our archives with the gzip option in the crawler, the average \format{} size shrinks down to 12 MB, driven largely by the compression of the ever-repeating XML format of the behavior records (11 MB to 1.5 MB).

% Overall, the additional storage needed for \format{} is less than double that of a HAR file recorded via a proxy. While the choice of format ultimately depends on the specific needs of each experiment, we believe the additional storage requirement is justified by the more detailed insights \format{} provides.
}

\subsection{Maintenance Overhead}
\label{sec:dis:maintenance}

\rev{\tool{} is developed as a series of modifications to ``stock'' Chromium, along with external tools to interact with those modifications. Maintaining modifications in a rapidly changing code base like Chromium is a challenge, which can lead to quickly-abandoned prototypes, especially in research. To address this \tool{} is designed as an ongoing solution, with design choices that dramatically reduce the maintenance overhead in keeping \tool{} current.
% \tool{} is intended to be an ongoing solution to the research and measurement problems discussed in prior sections, and in this section we briefly discuss some of the design choices that dramatically reduce the maintenance overhead in keeping \tool{} current.
}

\rev{Most of \tool{}'s complexity comes from modifications to Chromium and its sub-projects (\EG{} Blink, V8), which are mostly implemented in C++. The frequent changes to the code base make correctly maintaining modifications outside the project difficult. For example, patches that work correctly against one version of Chromium can need significant reworking for the next version, if Chromium developers have refactored code or added significant new functionality.}

\rev{\tool{} is implemented using strategies that simplify updates to new Chromium versions, and minimize amount of patching needed. First, most of \tool{}'s logic is implemented separately and independently of upstream Chromium code, kept in its own C++ namespace and data structures (instead of modifying the ones relied on by Chromium). Second, \tool{} leverages many of Chromium's systems to capture information about a page's execution without needing to modify Chromium code. For example, Chromium's \emph{CoreProbe} system allows code to receive notifications when certain page activities occur (\EG{} new elements created).
% Similarly, Chromium's ``mojo'' and ``devtools'' systems provide different mechanisms for inter-process communication, allowing \tool{} to work ``with the grain'' of Chromium's multi-process architecture.
Third, Chromium's build system and C++ classes include mechanisms to augment Chromium's classes in a more robust way than sub-classing. For example, the classes \texttt{Supplementable} and \texttt{Supplement} allow developers to add behavior to existing ``Supplementable'' classes, by implementing that new functionality in a ``peer Supplement'' class, without needing subclasses or changing upstream code, further making \tool{}'s behavior robust across Chromium changes. And fourth, when the previously discussed approaches are not possible, \tool{} uses subclasses when possible, and modifies Chromium code through clever use of C/C++'s pre-processor (instead of through patches), both of which, while not ideal, end up being easier to carry across Chromium versions than patches. All this allows us to maintain \tool{} with only 33 patches (compared with over 12k lines to implement \tool{}'s browser-side functionality.}

\rev{This design has been very successful in making \tool{} easy to maintain across Chromium releases. \tool{} has been kept current with every Chromium version since September 2022, (\IE{} 40 Chromium versions) with little to no changes required each update. We estimate less than an hour of time has been needed to keep \tool{} current for each Chromium release, many needing no additional effort at all.}

% the median amount of time being ``zero hours'', or 

% \rev{Like with any developed tool, a key question is who will maintain it and how costly is extending it. Part of our team is from \censor{Brave add random length} which actively integrates \tool{}'s functionalities into its \censor{pagegraph-crawl}.
% They have been tracking the necessary changes through use of established C++ patterns and Chromium constructs since \censor{XXXX}, reducing patch requirements to only XXX hours per patch. 
% To extend the functionality, a researcher would either need to modify the patches to Chromium or modify the \censor{pagegraph-crawl}, depending on the needs.
% TODO: Concrete in numbers, we talk about X lines of added code. Also talk about engineering costs with trade-offs and implications.
% }
% \section{A PageGraph Archive}

% We publish our archive for everyone.

\section{Conclusion}

This paper highlights a significant issue in the current \Web{} measurement landscape: the absence of general-purpose measurement tools and archives. Instead, we see self-implemented, single-use crawlers. That this approach easily leads to correctness and reproducibility issues is shown by a case study where we identified errors in a published paper misattributing event handlers. For \Web{} archives, we find over \empirical{13\%} discrepancy in \JS{} API call appearances between replayed archives and our baseline, revealing accuracy gaps.

% In conclusion, our paper highlights the problem that current Web measurement tools and archives are not sufficient enough to support reproducibility of experiments and proposes a new tool (\tool{}) and archiving format (\format{}) as solution. One problem, for example, our study revealed is the discrepancy of more than \empirical{13\%} in the appearance of \JS{} API calls between responses replayed from common Web archives and our baseline. Additionally, we identified issues in existing work leading to falsely attributed event handlers. This work clearly demonstrates how the current practice in the Web measurement field (\IE{} self-implemented crawlers and single-use crawls without archived data), leads to a lack of efficiency, correctness, and reproducibility of research.

To tackle these issues and achieve reproducible, efficient, and accurate \Web{} measurements, we then presented and evaluated \tool{}. It bundles a \Web{} page's HTTP communication, an execution graph, and screenshots into a \format{} archive, proving more accurate in replicating page behavior than traditional archives. Many existing measurement studies could have benefited from using it, offering a solution to current problems and improving future measurement research.

% Additionally, our literature analysis indicates that around \empirical{70\%} of surveyed papers could have used \tool{} as a crawler, and \empirical{48\%} could have queried a \format{} archive instead of running their own crawlers.
% We invite the community to create and share more datasets like ours to support future research. Additionally, we plan to provide a publicly available regularly-updated archive of \format{} files. We believe, future research can largely leverage on these datasets and archives, enhancing reproducibility, transparency and correctness in Web measurement research.

\section*{Acknowledgments}

We thank the USENIX reviewers for their valuable comments and suggestions, which contributed to improving the presentation of our paper. We are particularly grateful to our shepherd for their guidance and advice.
Additionally, we want to thank Jannis Rautenstrauch for his HTTPArchive skills, paper tips, and proofreading.

% Further, we want to thank several people who have done invaluable work on the \PG{} project at Brave Software, including Anton Lazarev (who did significant work maintaining earlier versions of \PG{}), Brian Johnson (who explained and designed many of the engineering strategies that make it possible to keep \PG{} up to date with Chromium), and in particular Aleksei Khoroshilov (who has done a tremendous amount of work on the project, including having \PG{} enabled by default in Brave Browser desktop builds). Lastly, we want to appreciate Michael Smith (University of California; San Diego) who did significant work on \PG{} during an internship at Brave.

Further, we want to thank several people who have done invaluable work on the \PG{} project at Brave Software, including Anton Lazarev, Brian Johnson, and, in particular, Aleksei Khoroshilov. Lastly, we want to appreciate the work on \PG{} done by Naomi Smith (University of California; San Diego) during an internship at Brave.

This work was conducted in the scope of
a dissertation at the Saarbrücken Graduate School of Computer
Science.
\section*{Ethical Considerations}
\label{apx:ethics}

As USENIX Security rightfully states, ethics plays a critical role in conducting research. Therefore, we carefully considered our project's ethical implications before we started our research. In our study, we sent GET requests to the 10k most popular origins according to the CrUX list~\cite{ruthWorldWideView2022}. This is a research practice employed by numerous \Web{} studies before and is generally considered ethical in the security and privacy community. However, it is essential to acknowledge that ethical standards can evolve, and each project should be assessed on a case-by-case basis.

In this study, we requested each origin of the top 10k only once per experiment with a standard GET request. Since a GET request is designed to receive data without making any modifications, we assume no negative impact for the end user. We also ensure no negative impact for the website operators as our minimal interaction does not contribute significantly to the common internet traffic or pose any more risk than typical internet background noise. While we, however, might potentially violate some websites' terms of service, as we did not verify whether automated site visits were explicitly allowed, we believe that the benefits to the research community in the form of our measurements and the demonstration of \tool{} outweigh this concern. All our additional measurements are performed on our client side, \IE{} our server, not influencing any third-party server. In conclusion, we are confident that our work did not cause harm to any individual or organization. Our research and measurements aim to contribute benefit to our research community by providing and influencing new ways how to conduct future \Web{} measurement studies.

\section*{Open Science}
\label{apx:availability} 
For future \Web{} measurement research, we keep our modified Chromium browser up to date and make it public for others to use~\cite{bravesoftwarePageGraph2023}. \tool{} is further actively developed as an open-source crawler under the name pagegraph-crawl~\cite{pg-crawl}. Additionally, we provide a query tool that allows querying the produced format~\cite{pg-query}. 
We plan to regularly update a collection of the top 10k websites collected in the \tool{} format.

To promote transparency and reproducibility in our research field, we publish the code we used in this paper along with this publication~\cite{experiment-pipeline}. This includes the pipeline to create our datasets and the scripts used to analyze the results. In addition, we have uploaded the top 1k entries of our datasets to Zenodo~\cite{experiment-dataset}. Due to its size, the full dataset is available upon request.
% \pagebreak

%-------------------------------------------------------------------------------

\bibliographystyle{plainurl}
\bibliography{references}

%\clearpage 
\appendix
\section{De-Obfuscated Google Analytics Code}
\label{apx:google-analytics}
Google Analytics uses programmatically added event listeners for an HTML node to react to events while still allowing CSP to block inline scripts (see~\Cref{lst:event-listener}).

\begin{listing}[h]
    \centering
        \begin{minted}[
            frame=lines,
            framesep=2mm,
            baselinestretch=1.2, 
            fontsize=\footnotesize,
            linenos,
            breaklines
        ]{javascript}
ele = document.createElement("script")
ele.type = "text/javascript"
ele.src = ff.createScriptURL(url)
ele.onload = loadHandler
ele.onerror = errorHandler
ele.setAttribute("nonce", nonce)
scr = document.getElementsByTagName("script")[0]
scr.parentNode.insertBefore(ele, scr)
        \end{minted}
    \caption{A deobfuscated example of Google Analytics code  programmatically setting event listeners.}
    \label{lst:event-listener}
\end{listing}

\section{Paper Collection Methodology}
\label{apx:papers}

We build the paper collection for \Cref{sec:availability} up on the SoK paper by Stafeev and Pellegrino \cite{stafeevSoKStateKrawlers2024}, who analyzed the use of crawlers in Web measurement studies. They include USENIX Security, S\&P, NDSS, CCS, as well as the specialized venues WWW, PETS, and IMC in their study, and collected all papers from 2010 to 2022 
that contain the keywords \texttt{tranco}, \texttt{alexa} in a combination with \texttt{top} or \texttt{site}.
From the resulting 1,057 papers, they removed all papers that did not employ automated crawling. The remaining papers build a list of 403 Web \revise{crawling} studies. Taking this list as a base, we analyzed all (137) papers from 2020 to 2022 to understand the individual uses of crawlers and whether or not \tool{} could have been used. We also only considered papers that measured data that are within the scope of what \tool{} offers, i.e., they either analyze webpage \JS{} executions, the page's content, or any HTTP requests made. This excludes studies using crawlers for other goals, such as network benchmarking purposes~\cite{meier2022ditto, pavur2021qpep} or side-channel attacks~\cite{tan2021invisible, zhan2022graphics} and narrowed our focus to \empirical{97} papers.

\end{document}